\documentstyle[11pt]{article}
\def\tsz{.9}

\def\se{\;\;=\;\;}

\def\/{\over}

\def\a{\alpha}
\def\b{\beta}
\def\c{\gamma}
\def\C{\Gamma}

\def\d{\delta}

\def\e{\epsilon}

\def\f{\phi}

\def\vf{\varphi}

\def\l{\lambda}
\def\L{\Lambda}
\def\m{\mu}
\def\n{\nu}

\def\r{\rho}
\def\s{\sigma}

\def\th{\theta}

\def\z{\zeta}
\def\beq{\begin{equation}}
\def\eeq{\end{equation}}
\def\beqa{\begin{eqnarray}}
\def\eeqa{\end{eqnarray}}
\def\barr{\begin{array}}
\def\earr{\end{array}}
\def\x{\xi}
\def\y{\eta}
\def\o{\omega}
\def\O{\Omega}
\def\del{\partial}

\def\ui{\underline {i}}
\def\bs{\nabla\!\!\!\!\!/}

\def\bn{\bar{\nabla}}

\let\la=\label

\let\bm=\bibitem

\def\nn{\nonumber}
\def\ed{\end{document}}
\def\ba{\begin{array}}
\def\ea{\end{array}}
\def\bea{\begin{eqnarray}}
\def\eea{\end{eqnarray}}
\def\ft#1#2{{\textstyle{{\scriptstyle #1} 
\over {\scriptstyle #2}}}}
\def\fft#1#2{{#1 \over #2}}
\def\wwb#1#2#3{$\left(\fft{#1}{#3},\,\fft{#2}{#3}\,\right)$}

\newcommand{\be}{\begin{equation}}
\newcommand{\ee}{\end{equation}}
\newcommand{\eq}[1]{(\ref{#1})}

\newcommand{\mx}[4]{\left#1\begin{array}{#2}#3\end{array}\right#4}
\newcommand{\w}[1]{\\[0.#1cm]}

\newcommand{\hoch}[1]{$\, ^{#1}$}
\newcommand{\lge}[1]{\ell\geq #1}

\begin{document}
\pagestyle{empty}
\rightline{CTP TAMU-15/98}
\rightline{hep-th/9804166}
\vspace{1.5truecm}
\centerline{\Large \bf Spectrum of $D=6$, $N=4b$ Supergravity on $AdS_{3}\times S^{3}$}
\vspace{1truecm}
\centerline{S. Deger\hoch{1}, A. Kaya\hoch{1}, 
E. Sezgin\hoch{2} and P. Sundell\hoch{3}}
\bigskip
\centerline{\it Center for Theoretical Physics, Texas A \& M University,}
\centerline{\it College Station, Texas 77843, USA}
\vspace{1.3truecm}
\centerline{ABSTRACT}
\vspace{.5truecm}

The complete spectrum of $D=6$, $N=4b$ supergravity with $n$ tensor
multiplets compactified on $AdS_{3}\times S^{3}$ is determined. The
$D=6$ theory obtained from the $K_3$ compactification of Type IIB string
requires that $n=21$, but we let $n$ be arbitrary. The superalgebra that
underlies the symmetry of the resulting supergravity theory in $AdS_{3}$
coupled to matter is $SU(1,1|2)_{L}\times SU(1,1|2)_{R}$. The theory
also has an unbroken global $SO(4)_R\times SO(n)$ symmetry inherited
from $D=6$. The\ \ spectrum of states arranges itself into a tower of
spin $2$ supermultiplets, a tower of spin $1$, $SO(n)$ singlet
supermultiplets, a tower of spin $1$ supermultiplets in the vector
representation of $SO(n)$ and a special spin $\ft12$ supermultiplet also
in the vector representation of $SO(n)$. The $SU(2)_{L}\times SU(2)_{R}$
Yang-Mills states reside in the second level of the spin $2$
tower and the lowest level of the spin $1$, $SO(n)$ singlet tower and
the associated field theory exhibits interesting properties.

\vspace{2cm}

{\vfill\leftline{}\vfill
\vskip	10pt
\footnoterule
{\footnotesize
\hoch{1}
Research supported in part by The Scientific and Technical Research
Council of Turkey (TUBITAK)\\
\hoch{2} Research supported in part by 
NSF Grant PHY-9722090 \\
\hoch{3} Research supported by The 
Natural Science Research Council of Sweden
(NFR) \vskip -12pt } 

\vfill\eject
\pagestyle{plain}


\section{Introduction}


Compactifications of supergravity on anti de Sitter spacetimes have been
studied extensively in the past (see \cite{ss} for a survey). In
particular, the $ AdS_4\times S^7$ compactification of $D=11$
supergravity has been a subject of much study (see \cite{dnp} for a
review). AdS spacetimes have also played role in the study
of supermembrane compactifications \cite{mc1}. More recently, AdS
backgrounds have made a remarkable appearance in the context of
M-theory, thanks to the Maldacena conjecture \cite{jm} according to
which, the low energy dynamics of M-theory on AdS is captured by the
dynamics of a conformal field theory on the boundary of AdS. The
calculational framework for testing this conjecture has been spelled out
in \cite{w1,gkp}.

Much of the work done so far in this subject has focused on the $AdS_5
\times S^5$ compactification of type IIB supergravity. While this is a
very interesting model to explore, the fact that the boundary theory is
$N=4$ super Yang-Mills theory in $3+1$ dimensions makes some of the
relevant calculations rather difficult to carry out. On the other hand,
$AdS_3$ compactifications of M-theory involve boundary conformal field
theories in $1+1$ dimensions which are far more amenable to exact
calculations. Furthermore, $AdS_3$ has been shown to arise in the
horizon geometry limit \cite{dgt} of the self-dual string \cite{dl}, the
horizon geometry limit of black holes in Type II string theory
\cite{hyun} and intersecting branes \cite{s1,s2}. More recently, $AdS_3$
bulk/boundary correspondence has been studied in \cite{ms,m,ads1,ads2}
with interesting results. In particular, the $AdS_3\times S^3 \times
K_3$ compactification of type IIB supergravity was considered in
\cite{jm,ms}. The $K_3 $ compactification gives rise to $D=6, N=4b$
supergravity coupled to 21 tensor multiplets. The covariant field
equations for this model were constructed by Romans \cite{romans1}. (The
quartic fermion terms for this model were recently obtained in
\cite{ric}). A subsequent compactification on $AdS^3\times S^3$ yields a
spectrum of states some of which have already been found and the
corresponding operators in the $AdS_3$ boundary conformal field theory
have been identified in \cite{ms}.

The purpose of this paper is to determine the complete spectrum of $D=6,
N=4b$ supergravity coupled to $n$ tensor multiplets. For anomaly-freedom
one sets $n=21$ but the results are valid for any $n$. The complete
spectrum is given in Tables \ref{htt}, \ref{hot} and \ref{hotn} and the
supermultiplet structure is shown in Figures \ref{httf}, \ref{hotf} and
\ref{hotnf}, with the notation for representations defined in \eq{drs}.
These results are expected to shed further light on the nature of the
boundary/bulk duality in the context of $AdS_3$. 

The organization of the paper is as follows. In section 2 we describe
the model and linearize the field equations. In section 3 the bosonic
sector of the model is studied; harmonic expansions on $S^3$ are
performed, following the techniques of \cite{pvn}, in the de
Donder-Lorentz gauge and the linearized field equations are
diagonalized. In section 4, the fermionic sector is studied by taking
similar steps. In section 5, $AdS_3$ is analytically continued to a
sphere $S^3_E$ on which harmonic expansions are carried out, following
the procedures described in \cite{es1,es2}. Continuing back to $AdS_3$
we then determine the complete representation content of all the
physical states. In section 6, the supermultiplet structure of the
complete spectrum is determined. We summarize our results in the
concluding section where we also comment on the nature of the sector
involving the Yang-Mills vector fields originating from the $S^3$
compactification. Notations and conventions are provided in the
Appendix.


 \section{The $D=6$, $N=4b$ Supergravity Plus $n$ Tensor Multiplets}

\subsection{Field Content}


The equations of motion for $D=6, N=4b$ supergravity coupled to $n$
tensor multiplets were constructed in \cite{romans1}. The theory consists
of a pure supergravity multiplet containing a graviton, $4$ gravitinos
and $5$ self-dual tensor fields coupled to $n$ tensor multiplets
containing an anti-self-dual tensor field, $4$ fermions and $5$ scalars.
The field content is summarized in the following Table.

\begin{center}
\begin{tabular}{|l|cccccc|}
\hline
&&&&&&\\
Field&$g_{MN}$ & $\psi_{M}$ & $B_{MN}^{i}$ & $B_{MN}^{r}$ & 
$\chi^{r}$ & $\f^{ir}$\\
&&&&&&\\
$SO(5)$~&$1$&$4$&$5$&$1$&$4$&$5$\\
&&&&&&\\
$SO(n)$~&$1$&$1$&$1$&$n$&$n$&$n$\\
&&&&&&\\
\hline
\end{tabular}\la{d6}
\end{center}

where $M,N,\dots= 1,\dots,6$ are curved $6D$ indices,
$i,j,\dots=1,\dots,5$ are $SO(5)$ vector indices and $r,s,\dots=1,\dots
n$ are $SO(n)$ vector indices. The $SO(5)$ spinor indices carried by
$\psi_M$ and $\chi^r$ are suppressed. 

In the coupled theory the scalar sector constitute a sigma-model over
the coset space

\be
{SO(5,n)\/SO(5)\times SO(n)}\ .\la{coset}
\ee

The scalars parametrize a vielbein $\left(V_{I}{}^{i}\, ,\
V_{I}^{r}\right)$ obeying the $SO(5,n)$ relation

\be
V_{I}{}^{i}V_{J}{}^{i}-V_{I}{}^{r}V_{J}{}^{r}\se \y_{IJ}\ ,\la{VV}
\ee

where the index $I=1,\dots,5+n$ transforms under global $SO(5,n)$
transformations and is raised and lowered with the metric
$\eta_{IJ}=diag~(+++++--\cdots-)$. The indices $(i,r)$ are transforming
under composite local $SO(5)\times SO(n)$ transformations. Defining

\be
dV V^{-1}\se \left(\begin{array}{cc}
Q^{ij}&\sqrt{2}P^{is}
\w1
\sqrt{2}P^{jr} & Q^{rs}\end{array}\right)\ ,
\ee

$Q^{ij}$ and $Q^{rs}$ are the composite $SO(5)$ and $SO(n)$ connections
solvable in terms of the $5n$ physical scalars $\f^{ir}$ via the
Cartan-Maurer equation. The covariant derivatives in the scalar and the
fermionic sectors are given by

\bea
D_{M}P_{N}{}^{ir}&=&\nabla_{M}P_{N}{}^{ir}-
Q_{M}{}^{ij}P_{N}{}^{jr}-Q_{M}{}^{rs}P_{N}{}^{is}\ ,
\nn\w1
D_{M}\chi^{r}&=&
\nabla_{M}\chi^{r}-\ft14Q_{M}{}^{ij}\C^{ij}\chi^{r}-
Q_{M}{}^{rs}\chi^{s}\ ,\la{cChi}
\w1
D_{M}\psi_{N}&=&
\nabla_{M}\chi_{N}-\ft14Q_{M}{}^{ij}\C^{ij}\chi_{N}
\ ,\nn
\eea

where $\C^{i}$ are the $SO(5)$ $\C$-matrices and $\nabla_{M}$ denotes
the Lorentz covariant derivative. In the spin-$1$ sector the
supersymmetric couplings are effected via the modified $3$-form field
strengths 

\be
H^{i}\se G^{I}V_{I}{}^{i}\ ,\quad
H^{r}\se G^{I}V_{I}{}^{r}\ ,\la{mfs}
\ee
 
where the elementary field strengths $G^{I}_{MNP}$ obey the
globally $SO(5,n)$ invariant Bianchi identity

\be
dG^{I}\se 0\ ,\quad G^{I}\se dB^{I}\ .
\ee


\subsection{Field Equations and Supersymmetry Transformations}


In the bosonic sector, the field equations are the Einstein equation

\be
R_{MN}\se
H^{i}_{MPQ}H^{i}_{N}{}^{PQ}+H^{r}_{MPQ}H^{r}_{N}{}^{PQ}
+2P_{M}{}^{ir}P_{N}{}^{ir}\ ,\la{E}
\ee

the scalar equation

\be
D^{M}P_{M}{}^{ir}-\ft{\sqrt{2}}{3}H^{i~MNP}H_{MNP}^{r}\se 0
\ ,\la{S}
\ee

and the following $SO(5)\times SO(n)$ invariant Hodge-duality conditions
on the $3$-form field strengths

\be
H_{MNP}^{i}\se\ft1{3!} e_{MNPQRS}~H^{i\,QRS}\ ,\qquad
H_{MNP}^{r}\se-\ft1{3!}e_{MNPQRS}~H^{r\,QRS}\ .\la{dual}
\ee

In the fermionic sector the field equations are

\bea
&&\C^{MNP}D_{N}\psi_{P}-H^{i\,MNP}\C_{N}\C^{i}\psi_{P}
+\ft12 H^{r\,MNP}\C_{NP}\chi^{r}-
\ft1{\sqrt{2}}P_{N}{}^{ir}\C^{N}\C^{M}\C^{i}\chi^{r}\se 0\ ,
\w1
&&\C^{M}D_{M}\chi^{r}+\ft1{12}\C^{MNP}H_{MNP}^{i}\C^{i}
\chi^{r}-\ft12 H^{r\,MNP}\C_{MN}\psi_{P}
-\ft1{\sqrt{2}}P_{N}{}^{ir}\C^{M}\C^{N}\C^{i}\psi_{M}
\se 0\ .\nn
\eea

The supersymmetry transformations acting covariantly on the field
equations are 

\bea
\d e_{M}{}^{A}&=& \bar{\e}\C^{A}\psi_{M}\ ,\nn
\w1
\d \psi_{M}&=& D_{M}\e-\ft14H^{i}_{MNP}\C^{NP}\C^{i}
\e\ ,\nn
\w1
\d B_{MN}^{I} &=& -V^{I\,i}~\bar{\e}\,\C_{[M}
\C^{i}\psi_{N]}
+\ft12 V^{I\, r}~\bar{\e}\,\C_{MN}\chi^{r}\ ,\la{susy}
\w1
\d \chi^{r} &=&
\ft1{\sqrt{2}}\C^{M}P_{M}{}^{ir}\C^{i}\e
+\ft1{12}\C^{MNP}H^{r}_{MNP}\e\ ,\nn
\w1
\d V_{I}{}^{i}&=&\bar{\e}\,\C^{i}\chi^{r} V_{I}{}^{r}\ ,\nn
\w1
\d V_{I}{}^{r}&=&\bar{\e}\,\C^{i}\chi^{r} V_{I}{}^{i} \ ,\nn
\eea

where $A=1,\dots,6$ is an $SO(5,1)$ tangentspace index and $e_{M}{}^{A}$
is the sechsbein. The superalgebra closes on general coordinate, local
Lorentz, tensor gauge and composite local $SO(5)\times SO(n)$
transformations.


\subsection{The Vacuum Solution}


We shall study the compactification on the maximally supersymmetric
vacuum solution with the geometry of $AdS_{3}\times S^{3}$, with
curvature tensors given by

\bea
R_{\m\n\r\s}&=& -m^{2}(g_{\m\r}\,g_{\n\s}-g_{\n\r}\,g_{\m\s})\ ,\nn
\w1
R_{abcd}&=& m^{2}(g_{ac}\,g_{bd}-g_{bc}\,g_{ad})\ ,\la{csthree}
\eea

where $\m,\n,\dots=1,2,3$ are curved $AdS_{3}$ indices,
$a,b,\dots=1,2,3$ are curved $S^{3}$ indices and $g_{\m\n}$ and $g_{ab}$
are the metrics of $AdS_{3}$ and $S^{3}$ with radius $m^{-1}$. One of
the components of the self-dual field strength is singled out and put equal to
the Levi-Cevita tensors, while the remaining field strengths vanish
 
\bea
H^{i}_{\m\n\r}&=&m \,e_{\m\n\r}\,\d^{i}_{5}\ ,\qquad
H^{i}_{abc}\se m \,e_{abc}\,\d^{i}_{5}\ ,
\nn\w1
H^{r}_{MNP}&=&0\ .
\eea

This solution uses a relative orientation between the six-dimensional
space-time and the two three-dimensional factors where $e_{\m\n\r
abc}=e_{\m\n\r}e_{abc}$. The $SO(5,n)$ vielbein is taken to be constant.
By a global $SO(5,n)$ rotation it can be set equal to unity

\be
V_{I}{}^{i}\se\d^{i}_{I}\ ,\qquad V_{I}{}^{r}\se \d^{r}_{I}\ .
\ee

Finally we set the fermions equal to zero

\be
\chi_{r}\se 0\ ,\qquad \psi_{M}\se 0\ .\la{fzero}
\ee

Supersymmetry of the solution requires that $\d\psi_{M}=0$ and
$\d\chi=0$. The latter is trivially satisfied, while the first condition
yields the Killing spinor equations 

\bea
&&D_{\m}\e+\ft12 m\,\c_{\m}\C^{5}\e\se 0\ ,\nn
\w2
&&D_{a}\e-\ft{i}2 m\,\c_{a}\C^{5}\e\se 0\ ,\la{ksst}
\eea

where the spinor depend on both $AdS_{3}\times S^3$ coordinates and
carries indices of $SO(2,1)\times SO(3)$ that have been supressed (see
the appendix for full details). The integrability of these equations
follows from \eq{csthree}.


\subsection{Linearized Field Equations}


We parametrize the linearized fluctuations around a background 
with non-vanishing metric $\bar{g}_{MN}$, three-form field strength
$\bar{H}^{i}_{MNP}$ and constant vielbein as follows

\bea
g_{MN}&=&\bar{g}_{MN}+h_{MN}\ ,
\nn\w{15}
G^{I}_{MNP}&=&\bar{G}^{I}_{MNP}+g^{I}_{MNP}\ ,
\nn\w{15}
V_{I}{}^{i}&=&\d^{i}_{I}+\f^{ir}\d^{r}_{I}\ ,
\la{gmn}\w{15}
V_{I}{}^{r}&=& \d^{r}_{I}+\f^{ir}\d^{i}_{I}\ ,
\nn\w{15}
P_{M}^{ir}&=&\ft1{\sqrt{2}}\del_{M}\f^{ir}\ .
\eea

The composite $SO(5)\times SO(n)$ connection $Q$ is then quadratic in
scalar fluctuations. Note that in the background the flat and curved
$SO(5,n)$ indices are identified. Taking the background to obey the
classical field equations to zeroth order, the resulting linearized
field equations for the bosonic fluctuations are

\bea
&&-\ft12\bn^{2}h_{MN}+R_{(M}{}^{P}h_{N)P}+
R_{PMNQ}h^{PQ}+\bn_{(M}\bn^{P}h_{N)P}-\ft12\bn_{M}\bn_{N}
\left(\bar{g}^{PQ}h_{PQ}\right)
\nn\w1
&&\qquad\qquad \se 2 \bar{H}^{i}_{PQ(M}g^{i}_{N)}{}^{PQ}
-2\bar{H}^{i}_{MP}{}^{Q}\bar{H}^{i}_{N}{}^{PR}h_{QR}\ ,\la{linee}
\w2
&&\ft16\bar{e}_{MNP}{}^{QRS}g^{i}_{QRS}\se
g^{i}_{MNP}-3h_{[M}{}^{Q}\bar{H}^{i}_{NP]Q}
+\ft12 h^{Q}{}_{Q}\bar{H}^{i}_{MNP}\ ,\la{linasd}
\w2
&&\ft16\bar{e}_{MNP}{}^{QRS}g^{r}_{QRS}\se
-g^{r}_{MNP}-2\f^{ir}\bar{H}^{i}_{MNP}\ ,\la{linsd}
\w2
&&\bn^{2}\f^{ir}\se \ft23 g^{r,MNP}\bar{H}^{i}_{MNP}\ .\la{linse}
\eea

and

\bea
&&\C^{MNP}D_{N}\psi_{P}-{\bar H}^{i\,MNP}\C_{N}\C^{i}\psi_{P} \se 0 \ ,
\nn\w1
&&\C^{M}D_{M}\chi^{r}+\ft1{12}\C^{MNP}{\bar H}_{MNP}^{i}\C^{i}
\chi^{r} \se 0\ .\la{chie2}
\eea

for the fermionic fluctuations. The linearized field equations are
invariant under the linearized reparametrizations and tensor gauge
transformations

\bea
\d h_{MN}&=& \bn_{M}\x_{N}+\bn_{N}\x_{M}\ ,\nn
\w1
\d B^{I}_{MN}&=& 2\del_{[M}\y^{I}_{N]}+
\d^{I\,i}\x^{P}\bar{H}^{i}_{MNP}\ ,\la{lgtb}
\w1
\d \f^{ir}&=& 0\ .\nn
\eea

The general coordinate transformation part of \eq{lgtb} is the Lie
derivative of $B_{MN}^{I}$ combined with a tensor gauge transformation
with the parameter $-\x^{Q}B^{I}_{QP}$.


\section{The Bosonic Sector}



\subsection{Harmonic Expansion on $S^{3}$ and Gauge Fixing}


Around the $AdS_{3}\times S^{3}$ background we parametrize the
metric fluctuations as 

\bea
h_{\m\n}&=& H_{\m\n}+\bar{g}_{\m\n}M\ ,\qquad 
\bar{g}^{\m\n}H_{\m\n}\se 0\ ,\nn
\w1
h_{\m a}&=& K_{\m a}\ ,\la{mtsplit}
\w1
h_{ab}&=&L_{ab}+\bar{g}_{ab}N\ ,\qquad 
\bar{g}^{ab}L_{ab}\se 0\ ,\nn
\eea

and anti-symmetric tensor fluctuations as

\bea
g^{I}_{MNP}&=&3\del_{[M}b^{I}_{NP]}\ ,
\nn\w1
b^{I}_{\m\n}&=&e_{\m\n\r}X^{I\,\r}\ ,\qquad
b^{I}_{ab}\se e_{abc}U^{I\,c}\ ,\qquad
b^{I}_{\m a}\se Z_{\m a}^{I}\ ,\la{tfp}
\eea

where the metric and the Levi-Cevita tensors are understood to be in the
background geometry. From here on we shall supress the bar-notation on
any quantity defined with respect to the background geometry. 

Expanding the fluctuations in harmonic functions on $S^{3}$ one has

\bea
H_{\m\n}(x,y~)&=&\sum H_{\m\n}^{(\ell\,0)}(x)Y^{(\ell\,0)}(y~)\ ,\nn
\w1
M(x,y~)&=&\sum M^{(\ell\,0)}(x)Y^{(\ell\,0)}(y~)\ ,\nn
\w1
K_{\m a}(x,y~)&=&\sum\left[K_{\m}^{(\ell,\pm1)}(x)Y_{a}^{(\ell,\pm1)}(y~)
+K_{\m}^{(\ell\,0)}(x)\del_{a}Y^{(\ell\,0)}(y~)\right]\ ,\nn
\w1
L_{ab}(x,y~)&=&\sum \left[L^{(\ell,\pm2)}(x)Y^{(\ell,\pm2)}_{ab}(y~)
+L^{(\ell,\pm1)}(x)\nabla_{\{a}Y_{b\}}^{(\ell,\pm1)}(y~)
+L^{(\ell\,0)}(x)\nabla_{\{a}\nabla_{b\}}Y^{(\ell\,0)}(y~)\right]\ ,\nn
\w1
N(x,y~)&=&\sum N^{(\ell\,0)}(x)Y^{(\ell\,0)}(y~)\ ,\la{n}
\w1
X^{I}_{\m}(x,y~)&=&\sum X^{I(\ell\,0)}_{\m} (x) Y^{(\ell\,0)}(y~)\ ,\nn
\w1
Z_{\m a}^{I} (x,y~)&=&\sum \left[Z_{\m}^{I(\ell,\pm1)}(x) Y^{(\ell,\pm1)}_{a}(y~) +
Z_{\m}^{I(\ell\,0)}(x) \del_{a}Y^{(\ell\,0)}(y~) \right]\ ,\nn
\w1
U_{a}^{I}(x,y~)&=& \sum \left[U^{I(\ell,\pm1)}(x)Y^{(\ell,\pm1)}_{a}(y~)+
U^{I(\ell\,0)}(x)\del_{a}Y^{(\ell\,0)}(y~)\right]\ ,\nn
\w1
\f^{ir}(x,y~)&=&\sum \f^{ir(\ell\,0)}(x)Y^{(\ell\,0)}(y~)\ ,\nn
\eea

where $(x,y~)$ are the coordinates of $AdS_{3}\times S^{3}$ and $\{ab\}$
denotes the traceless symmetric part. The harmonic functions
$Y_{(s)}^{(\ell_1 \ell_2)}$, where $(s)$ denotes the $SO(3)$ content, are
obtained from the Wigner functions of $SO(4)$ in the irreducible
representation labelled by the highest weight vector
$(\ell_{1}\,\ell_{2}\,)$ satisfying the restriction

\be
Y_{ (s)}^{(\ell_1 \ell_2)}\ :\qquad \ell_{1}\,\geq\,|\ell_{2}|\ .
\la{rest}
\ee

The dimension of the representation with highest weight $(\ell_1\ell_2)$ is 

\be
d_{(\ell_1\ell_2)}= (\ell_1+1)^2-\ell_2^2\ .
\ee

The action of the d'Alembertian on the Wigner functions is determined by
group theoretical means to be \cite{ss2}

\be
\nabla^{2}_{\,y}~Y_{ (s)}^{(\ell_1 \ell_2)} = -\left[\ell_1(\ell_1+2) + 
\ell_2^2 -s(s+1)\right]\,Y_{ (s)}^{(\ell_1 \ell_2)}\ ,\la{dely}
\ee

where {\it we have set the $S^3$ radius equal to one}. The right hand
side is simply the difference between the second order Casimir of
$SO(4)$ in the $(\ell_1,\ell_2)$ representation and that of $SO(3)$ in
the spin $s$ representation. Applying this formula to the different
cases yields 
  
\bea
\nabla^{2}_{\,y} Y^{(\ell\,0)}&=& \left[1-(\ell+1)^{2}\,\right] Y^{(\ell\,0)}\ ,\nn
\w2
\nabla^{2}_{\,y} Y^{(\ell,\pm1)}_{a}&=& \left[2-(\ell+1)^{2}\,\right] Y_a^{(\ell,\pm1)}\ , 
\quad\quad \nabla^{a}Y_{a}^{(\ell,\pm1)} \se  0\ , \la{yn1}
\w2
\nabla^{2}_{\,y} Y^{(\ell,\pm2)}_{ab}&=& \left[3-(\ell+1)^{2}\,\right] 
Y^{(\ell,\pm2)}_{ab}\ ,
\quad\quad \nabla^{a}Y_{ab}^{(\ell,\pm2)} \se 0\ ,\quad\quad 
\bar{g}^{ab}Y^{(\ell,\pm2)}_{ab} \se 0\ ,\nn
\w2
e_{a}{}^{bc}\del_{b}Y_{c}^{(\ell\,,~\pm1)}
&=& \pm(\ell+1)Y^{(\ell\,,~\pm1)}_{a} \ .  \nn
\eea

The harmonic one-forms $Y_{a}^{(\ell,\pm1)}$ are transverse, and
together with the longitudinal modes $\del_{a}Y^{(\ell\,0)}$ they form
complete set of one-forms $S^{3}$. Similarly the transverse
$Y_{ab}^{(\ell,\pm2)}$ modes and the longitudinal modes
$\nabla_{\{a}Y_{b\}}^{(\ell,\pm1)}$ and
$\nabla_{\{a}\nabla_{b\}}Y^{(\ell\,0)}$ span the space of traceless and
symmetric tensors on $S^{3}$. Note the zero modes in the form of the  
single constant mode $Y^{(00)}$, the six Killing vectors $Y_{a}^{(1,\pm1)}$ in
the adjoint representation of $SO(4)$ and the four conformal Killing
vectors $\del_{a}Y^{(10)}$. These obey

\be
\del_{a}Y^{(00)}\se0\ ,\qquad \nabla_{\{a}\nabla_{b\}}Y^{(10)}\se 0\ ,\qquad
\nabla_{(a}Y^{(1,\,\pm1)}_{b)}\se0\ .\la{zm}
\ee

The Killing vectors and the conformal Killing vectors together generate
the conformal group $SO(4,1)$ of the internal $S^3$.

By a field dependent gauge transformation generated by the harmonic
non-zero modes, we can remove all longitudinal gauge modes:

\bea
&&\ell\geq 1\ :\qquad 
K^{(\ell\,0)}_{\m}\se  Z_{\m}^{I(\ell\,0)}\se 
U^{I(\ell,\pm1)}\se0
\ ,\nn\w2
&&\ell\geq2\ :\qquad L^{(\ell,\pm1)}\se L^{(\ell\,0)}\se0\ .\la{bgm}
\eea

This shows the admissibility of the so called de Donder - Lorentz gauge 

\be
\nabla^{a}h_{\{ab\}}\se0\ ,\qquad
\nabla^{a}h_{a\m}\se0\ ,\qquad
\nabla^{a}b^{I}_{a M}\se0\ ,\la{ddl}
\ee

where the harmonic expansions \eq{n} thus read 

\be
\begin{array}{lclclcl}
H_{\m\n}&=&\sum H_{\m\n}^{(\ell\,0)}Y^{(\ell\,0)}&\quad,\qquad&
K_{\m a}&=&\sum K_{\m}^{(\ell,\pm1)}Y_{a}^{(\ell,\pm1)}\ ,\w1
M&=& \sum M^{(\ell\,0)}Y^{(\ell\,0)}&\quad,\qquad&
X^{I}_{\m}&=&\sum X_{\m}^{I(\ell\,0)}Y^{(\ell\,0)}\ ,\w1
L_{ab}&=&\sum L^{(\ell,\pm2)}Y^{(\ell,\pm2)}_{ab}&\quad,\qquad&
Z^{I}_{\m a}&=&\sum Z^{I(\ell,\pm1)}_{\m}Y^{(\ell,\pm1)}_{a}\ ,\w1
N &=&\sum N^{(\ell\,0)}Y^{(\ell\,0)}&\quad,\qquad&
U^{I}_{a}&=& \sum U^{I(\ell\,0)}\del_{a}Y^{(\ell\,0)}\ .
\end{array}
\la{gfmf}
\ee


\subsection{$AdS_{3}$ Gauge Symmetries}


The de Donder - Lorentz gauge completely fix the gauge symmetries that
are spontaneously broken by the compactification. These symmetries are generated by
harmonic non-zero modes and they transform states of different energy
into each other. The de Donder - Lorentz gauge does not touch the
remaining unbroken gauge symmetries that are generated by the the
harmonic zero modes and that correspond to local gauge symmetries in
$AdS_{3}$. The gauge fixing of these symmetries involves fixing of residual gauge
transformations leading to decoupling of longitudinal modes from the
physical spectrum and they therefore need special treatment. The gauge
symmetries are:

\begin{itemize}

\item The Stueckelberg shift symmetries 

\bea
\d H_{\m\n}^{(10)}(x)&=&-2\nabla_{\{\m}\nabla_{\n\}}\l^{(10)}(x)
\ ,\nn
\w1
\d M^{(10)}(x)&=&-\ft23\nabla^{2}_{\,x}\l^{(10)}(x)
\ ,\qquad
\d N^{(10)}(x)\se-2\l^{(10)}(x)\ ,\la{dN}
\w1
\d X^{I(10)}_{\m}(x)&=&-\d^{I\,5}\del_{\m}\l^{(10)}(x)
\ ,\qquad
\d U^{I(10)}(x)\se\d^{I\,5}\l^{(10)}(x)\ .\nn
\eea

\la{ck}

\item The $AdS_{3}$ reparametrizations and tensor gauge
transformations 

\bea
\d M^{(00)}(x)&=&\ft23 \nabla^{\m}\x_{\m}(x)
\ ,\qquad
\d H_{\m\n}^{(00)}(x)\se2\nabla_{\{\m}\x_{\n\}}(x)
\ ,\nn\w1
\d X^{I(00)}_{\m}(x)&=&-e_{\m}{}^{\n\r}\del_{\n}\y^{I}_{\r}(x)
+\d^{I\,5}\x_{\m}(x)\ ,\qquad \d N^{(00)}(x)\se0\ .\la{diff}
\eea

\item The $SO(4)$ Yang-Mills symmetries

\bea
\d K_{\m}^{(1\,,\,\pm1)}(x)&=&\del_{\m}\L^{(1\,,\,\pm1)}(x)\ ,
\nn\w1
\d Z_{\m}^{I(1\,,\,\pm1)}(x)&=&\mp~ \d^{I\,5}\ft{1}{2}~
\del_{\m}\L^{(1\,,\,\pm1)}(x)\ .\la{ymt2}
\eea

\end{itemize}

The Lorentz gauge \eq{ddl} for the anti-symmetric tensor fluctuations
allows no residual gauge transformations, since there are no harmonic
one-forms on $S^3$. For the same reason, the Yang-Mills transformation
\eq{ymt2} contains a compensating $y$-dependent tensor gauge
transformation of $b^{5}_{ab}$ in order to preserve the Lorentz gauge
condition (that is why $\d Z_{\m}^{5(1,\,\pm1)}$ is non-zero
in \eq{ymt2}).


{\tiny
\begin{table}[h]
\begin{center}
\begin{tabular}{|lll|c|c|c|c|c|c|c|}
\hline
&&&&&&&&&\\
&&Unbroken&&&&&&&\\
Sector:&&symmetry:&$h_{\m\n}$&$h_{\m a}$&$h_{ab}$&$B^{I}_{\m\n}$&
$B^{I}_{\m a}$&$B^{I}_{ab}$&$\f^{ir}$\\
&&&&&&&&&\\
\hline
&&&&&&&&&\\
$(\ell\,0)\,,\ \ell\geq2$&&-&$H_{\m\n}\,,\ M$&&$N$&$X_{\m}^{5}$&&
$U^{5}$&\\
&&&&&&&&&\\
$(10)$&&$\l$&$H_{\m\n}\,,\ M$&&$N$&
$X_{\m}^{5}$&&$U^{5}$&\\
&&&&&&&&&\\
$(00)$&&$\x_{\m}\,,\ \y^{5}_{\m}$&$H_{\m\n}\,,\ M$&&$N$&
$X_{\m}^{5}$&&&\\
&&&&&&&&&\\
$(\ell\,0)\,,\ \ell\geq1$&&-&&&&$X_{\m}^{\ui}$&& $U^{\ui}$&\\
&&&&&&&&&\\
$(00)$&&$\y^{\ui}_{\m}$&&&& $X_{\m}^{\ui}$&&&\\
&&&&&&&&&\\
$(\ell\,0)\,,\ \ell\geq1$&&-&&&&$X^{r}_{\m}$&& $U^{r}$&$\f^{5r}$\\
&&&&&&&&&\\
$(00)$&&$\y^{r}_{\m}$&&&& $X_{\m}^{r}$&&&$\f^{5r}$\\
&&&&&&&&&\\
$(\ell\,0)\,,\ \ell\geq0$&&-&&&&&&&$\f^{\ui,\,r}$\\
&&&&&&&&&\\
\hline
&&&&&&&&&\\
$(\ell\,,\,\pm1)\,,\ \ell\geq2$&&-&&$K_{\m}$&&&$Z^{5}_{\m}$ &&\\
&&&&&&&&&\\
$(1\,,\,\pm1)$&&$\L$&&$K_{\m}$&&&$Z^{5}_{\m}$ &&\\
&&&&&&&&&\\
$(\ell,\,\pm1)\,,\ \ell\geq1$&&-&&&&& $Z_{\m}^{\ui}$&&\\
&&&&&&&&&\\
$(\ell,\,\pm1)\,,\ \ell\geq1$&&-&&&&& $Z_{\m}^{r}$&&\\
&&&&&&&&&\\
\hline
&&&&&&&&&\\
$(\ell2)\,,\ \ell\geq2$&&-&&&$L$&&&&\\
&&&&&&&&&\\
\hline
\end{tabular}
\end{center}
\caption{{\small The $SO(4)$ representations arising in the harmonic expansion
of the bosonic fields in the de Donder - Lorentz gauge. The fields in
each row are described by a coupled system of equations which must be
diagonilized to find the spectrum. The residual gauge symmetries are
indicated by their parameters. The index $\ui$ labels the vector
representation of the unbroken $SO(4)_{R}\subset SO(5)$.}}

\la{tab1}
\end{table}
}


\subsection{ The Linearized Field Equations in the de Donder - Lorentz
Gauge}


In the de Donder - Lorentz gauge the linearized Einstein equations take
the form

\bea
&&-(\nabla^{2}_{\,x}+\nabla^{2}_{\,y}+2)H_{\m\n}+2\nabla_{\{\m}\nabla^{\r}H_{\n\}\r}
-\nabla_{\{\m}\nabla_{\n\}}(M+3N)\se0\ ,
\la{ee1}\w2
&&-(\nabla_{x}^{2}+\ft34\nabla^{2}_{y}+6)M
+\ft12\nabla^{\m}\nabla^{\n}H_{\m\n}-
\ft34\nabla^{2}_{\,x} N+6\nabla^{\m}X^{5}_{\m}\se0\ ,
\la{ee2}\w2
&&-(\nabla^{2}_{\,x}+\nabla^{2}_{\,y})K_{\m a}+\nabla_{\m}\nabla^{\n}K_{\n a}
+\nabla_{a}\nabla^{\n}H_{\m\n}-2\nabla_{\m}\nabla_{a}(M+N)\nn
\w2
&&\qquad\qquad+~4\nabla_{a}X_{\m}^{5}-4\nabla_{\m}U_{a}^{5}-
4e_{\m}{}^{\n\r}\nabla_{\n}Z^{5}_{\r a}+
4e_{a}{}^{bc}\nabla_{b}Z^{5}_{\m c}\se 0\ ,
\la{ee3}\w2
&&-\left(\nabla^{2}_{\,x}+\nabla^{2}_{\,y}-2\right)L_{ab}+2\nabla_{\{a}\nabla^{\m}K_{|\m|b\}}
-\nabla_{\{a}\nabla_{b\}}\left(3M+N\right)\se0\ ,
\la{ee4}\w2
&&-\left(\nabla_{x}^{2}+\ft43\nabla^{2}_{y}-8\right)N-
\nabla^{2}_{y}M-8\nabla^{a}U^{5}_{a}\se0\ ,\la{ee5}
\eea

where $\nabla_{x}^{2}=\nabla^{\m}\nabla_{\m}$ and
$\nabla_{y}^{2}=\nabla^{a}\nabla_{a}$. Each one of the linearized
Hodge-duality conditions \eq{linasd} and \eq{linsd} splits into two
pairs of equations, for the indices $(\m\n\r,abc)$ and $(\m\n c,\m bc)$,
where the two equations in each pair turn out to be equivalent. The
result is

\bea
&&\nabla^{\m}X^{i}_{\m}-\nabla^{a}U^{i}_{a}+\ft32\d^{i5}\left(
N-M\right)\se 0\ ,
\la{sd1}\w2
&&\nabla_{a}X^{i}_{\m}+\nabla_{\m}U^{i}_{a}-
e_{\m}{}^{\n\r}\,\nabla_{\n}Z^{i}_{\r a}-e_{a}{}^{bc}\,\nabla_{b}Z^{i}_{\m c}
-\d^{i5}K_{\m a}\se0\ ,
\la{sd2}\w2
&&\nabla^{\m}X^{r}_{\m}+\nabla^{a}U^{r}_{a}+2\f^{5r}\se0\ ,
\la{sd3}\w2
&&\nabla_{a}X^{r}_{\m}-\nabla_{\m}U^{r}_{a}-
e_{\m}{}^{\n\r}\nabla_{\n}Z^{r}_{\r a}
+e_{a}{}^{bc}\,\nabla_{b}Z^{r}_{\m c}\se0\ .\la{sd4}
\eea

Finally, the scalar equation \eq{linse} in the $AdS_{3}\times S^{3}$
vacuum reads

\be
-(\nabla^{2}_{\,x}+\nabla^{2}_{\,y})\f^{ir}-4\d^{i5}\left(
\nabla^{\m}X^{r}_{\m}-\nabla^{a}U^{r}_{a}\right)\se 0\ .\la{sce}
\ee

We next insert the harmonic expansions into the above equations. This
leads to the following set of irreducible equations (see Table \ref{tab1}
for the grouping of the fields according to their $SO(4)$ content):

\begin{itemize} 


\item {\bf The $J=2$ sector}


\bigskip

$(\ell\,0)$, $\ell\geq2$: $H_{\m\n}^{(\ell\,0)}$ gives rise to a massive
spin-$2$ mode described by the transverse and symmetric spin-$2$ tensor

\bea
&&S^{(\ell\,0)}_{\m\n}\se H^{(\ell\,0)}_{\m\n}-
\ft2{(\ell+1)^{2}}\nabla_{\{\m}\nabla_{\n\}}\left(
N^{(\ell\,0)}-2U^{(\ell\,0)}\right)\ ,
\nn\w2
&&\nabla^{\m}S^{(\ell\,0)}_{\m\n}\se0\ ,\la{mj2}
\eea

satisfying the equation

\be 
\left[\nabla^{2}_{\,x} +3-(\ell+1)^{2}\,\right]S^{(\ell\,0)}_{\m\n}\se0\ .\la{ste}
\ee

In obtaining this equation, we have used 

\bea
&&
\nabla^{\n}H_{\m\n}^{(\ell\,0)}-2\del_{\m} M^{(\ell\,0)}
-2\del_{\m}N^{(\ell\,0)}+4 X_{\m}^{5(\ell\,0)}-4\del_{\m}U^{5(\ell\,0)}
\se0\ , \quad \lge1\ ,
\eea

which follows from \eq{ee3}. We have also used \eq{yns} and

\be
3M^{(\ell\,0)}+N^{(\ell\,0)} \se0\ , \quad \lge2 .\la{mn}
\ee

\bigskip


\item  {\bf The $J=1$ sector} 


\bigskip

The relevant vectors are contained in the $(\ell, \pm1)$ sectors, which
we group as follows:

\begin{enumerate}

\item[$i$)] $(\ell,\pm1)$, From \eq{ee3} and \eq{sd2} we find that 
the two vectors $\ell\geq2$: $K_{\m}^{(\ell,\pm1)}$ and
$Z^{5(\ell,\pm1)}_{\m}$ are described by the coupled system

\bea
&&
\left( \nabla^{2}_{\,x} +\left[2-(\ell+1)^2\right] \right)
\,K_{\m}^{(\ell,\pm1)}-\nabla_{\m}
\nabla^{\n}K_{\n}^{(\ell,\pm1)}\ 
\nn\w2
&& 
\qquad\quad\  +4e_{\m}{}^{\n\r}\del_{\n}Z^{5(\ell,\pm1)}_{\r}
\mp\,4(\ell+1)Z_{\m}^{5(\ell,\pm1)} \se 0\ , 
\la{kzfb}\w2
&&e_{\m}{}^{\n\r}\del_{\n}Z^{5\,(\ell,\pm1)}_{\r}
\pm(\ell+1)Z^{5\,(\ell,\pm1)}_{\m} +K_\m ^{(1,\,\pm1)}\se 0 \ , \la{kzfa}
\eea

subject to the condition

\be
\nabla^{\m}K_{\m}^{(\ell,\pm1)}\se0\ , \quad \lge2\ ,\la{lgc}
\ee

which follows from \eq{ee4}.

\item[$ii)$] $(\ell,\pm1)$, $\ell\geq1$: From \eq{sd2} we find that the
vectors $Z^{\ui\,(\ell,\pm1)}_{\m}$ 
are described by the equations

\be
e_{\m}{}^{\n\r}\del_{\n}Z^{\ui\,(\ell,\pm1)}_{\r}
\pm(\ell+1)Z^{\ui\,(\ell,\pm1)}_{\m} \se0 \ ,\la{zui}
\ee

where we have introduced the index $\ui=1,...,4$, which labels the
vector representation of the unbroken $SO(4)_{R}\subset SO(5)$.

\item[$iii$)] $(\ell,\pm1)$, $\ell\geq1$: From \eq{sd4} we find that the vectors
$Z^{r(\ell,\pm1)}_{\m}$ 
are described by the equations

\be
e_{\m}{}^{\n\r}\del_{\n}Z^{r(\ell,\pm1)}_{\r}
\mp (\ell+1)Z^{r(\ell,\pm1)}_{\m}\se0\ \ .\la{zr}
\ee

\end{enumerate}

The the first order equations automatically imply the Lorentz condition
and when squared they yield the usual second order Proca equations for
massive vector fields. The first order nature of the equations however
implies that only half the number of helicity states are found in the
spectrum compared to the case of a second order Proca equation. The
vector spectrum will be analyzed in more detail in section 5 using
harmonic expansion on $AdS_{3}$.

\bigskip


\item  {\bf The $J=0$ sector}


\bigskip

Using the algebraic equation \eq{mn} and the $S^3$ harmonic expansion
formulae, yields four sets of coupled equations (see Table
1) that are diagonilized as follows:

\begin{enumerate}

\item[$i$)] $(\ell\,0)$, $\ell\geq2$: $M^{(\ell\,0)}$, $N^{(\ell\,0)}$,
$U^{5(\ell\,0)}$ and $X^{5(\ell\,0)}$ yields two scalar modes and a
massive spin $2$ mode. Analyzing \eq{ee2}, \eq{ee5} and \eq{mn} we find that
the two scalar modes are the two eigenmodes of

\bea
&&\nabla^{2}_{\,x} \mx{(}{c}{U^{5(\ell\,0)}\\N^{(\ell\,0)}}{)}-
\mx{[}{ll}{\ell(\ell+2) &2\\8\ell(\ell+2) & \ell(\ell+2)+8}{]}
\mx{(}{c}{U^{5(\ell\,0)}
\la{yns}\w2
N^{(\ell\,0)}}{)}\se 0\ .
\eea

This equation system is diagonalized as

\be
\left[ \nabla^{2}_{\,x} -\ell(\ell+2)-4 \pm 4(\ell+1)\,\right]N_{\pm}^{(\ell\,0)}=0\ .
\la{ynpm}
\ee

The vector $X^{5(\ell\,0)}_{\m}$ is not an independent field and from \eq{sd2}
we find that it is
given by 

\be
X^{5(\ell\,0)}_{\m}\se-\del_{\m}U^{5(\ell\,0)}\ . \la{men}
\ee

\item[$ii$)] $(\ell,\pm2)$, $\ell\geq 2$: From \eq{ee4} we find that
$L^{(\ell,\pm2)}$ describes
the scalar modes 

\be
(\nabla^{2}_{\,x}+1-(\ell+1)^{2})L^{(\ell,\,\pm2)}\se0\ \la{lpt}
\ee

\item[$ii$)] $(\ell\,0)$, $\ell\geq 1$: From \eq{sd1} and \eq{sd2} follows 
that $X^{\ui (\ell\,0)}_{\m}$ and
$U^{\ui (\ell\,0)}$ yield four scalar modes

\be
\left[\nabla^{2}_{\,x} +1-(\ell+1)^{2}\,\right] U^{\ui (\ell\,0)}\se 0\ ,
\qquad X_{\m}^{\ui (\ell\,0)}\se -\del_{\m}U^{\ui(\ell\,0)}\ ,\la{fsm}
\ee

\item[$iv$)] $(\ell\,0)$, $\ell\geq1$: From \eq{sd3} and \eq{sce}
follows that $X_{\m}^{r (\ell\,0)}$,
$U^{r(\ell\,0)}$ and $\f^{5r(\ell\,0)}$ yield $2n$ scalar modes
described by the eigenmodes of 

\be
\nabla^{2}_{\,x}\mx{(}{c}{U^{r(\ell\,0)}\\ \f^{5r(\ell\,0)}}{)}
-\mx{[}{ll}{\ell(\ell+2)& -2\\
-8\ell(\ell+2) &\ell(\ell+2)+8}{]}
\mx{(}{c}{U^{r(\ell\,0)}\\ \f^{5r(\ell\,0)}}{)}\se0\ ,\la{yf}
\ee

which is diagonalized exactly as in \eq{ynpm}:

\be
\left[ \nabla^{2}_{\,x} -\ell(\ell+2)-4 \pm 4(\ell+1)\,\right]
\phi_{\pm}^{r\,(\ell\,0)}=0\ . \la{ynpm2}
\ee

From \eq{sd4} follows that the vector $X_{\m}^{(\ell\,0)}$ is given by 			

\be
X_{\m}^{r(\ell\,0)}\se \del_{\m}U^{r(\ell\,0)}\ .
\ee

\item[$v$)] $(\ell\,0)$, $\ell\geq0$: From \eq{sce} follows that
$\f^{\,\ui r(\ell\,0)}$ yields $4n$ scalar modes which from \eq{sce} 

\be
\left[\nabla^{2}_{\,x} +1-(\ell+1)^{2}\,\right] 
\f^{\,\ui r (\ell\,0)}\se0\ .\la{fosm}
\ee

\end{enumerate}

\end{itemize}

There remains the analysis of the zero mode sectors transforming under
the $AdS_{3}$ gauge symmetries discussed in section 3.2 
(see Table \ref{tab1}).
These sectors are:

\begin{itemize}


\item {\bf The $(10)$ Sector}


The Stueckelberg shift symmetry \eq{dN} acts on the spin $2$ mode
$H_{\m\n}^{(10)}$ and the four scalar modes $M^{(10)}$, $N^{(10)}$,
$X_{\m}^{5(10)}$ and $U^{5(10)}$. For $\ell=1$ the algebraic condition
\eq{mn} does not follow from Einstein equations of motion due to the
condition on the conformal Killing vectors $\del_{a}Y^{(10)}$ given in
\eq{zm}. However, using a gauge transformation \eq{dN} with shift parameter
$\l^{(10)}$ obeying

\be
(\nabla^{2}_{\,x} +1)\l^{(10)}\se \ft12\left(3M^{(10)}+N^{(10)} 
\right)\ ,
\ee

we may instead impose \eq{mn} as the gauge condition

\be
3M^{(10)}+N^{(10)}\se0\ .\la{mng}
\ee

This gauge choice implies \eq{yns}, where $\ell$ is now to be set equal
to $1$, and a massive spin $2$ mode described by \eq{mj2} for $\ell=1$,
that is

\bea
&&S^{(1\,0)}_{\m\n}\se H^{(1\,0)}_{\m\n}-
\ft12\nabla_{\{\m}\nabla_{\n\}}\left(
N^{(1\,0)}-2U^{(1\,0)}\right)\ ,\qquad\nabla^{\m}S^{(1\,0)}_{\m\n}\se0
\ ,\nn\w2
&&\left(\nabla^{2}_{\,x} -1\,\right)S^{(\ell\,0)}_{\m\n}\se0\ .\la{mstm}
\eea

We have to remember, though, that the gauge choice \eq{mng} allows
residual shift transformations with parameter $\l^{(10)}$ satisfying
\be
(\nabla^{2}_{\,x} +1)\l^{(10)}\se 0\ .\la{leq}
\ee

This symmetry can be used to gauge away the scalar mode $N^{(10)}_{+}$.
We are thus left with the scalar mode

\be
(\nabla^{2}_{\,x} -15)N_{-}^{(1\,0)}=0\ .\la{wat}
\ee

Finally, the vector is given by

\be
X_{\m}^{5(10)}\se -\del_{\m}U^{5(10)}\ .
\ee


\item {\bf The $(00)$ Sector}


This sector can be grouped as follows (see Table \ref{tab1}):

\begin{enumerate}

\item[$i$)] The only contribution to the spectrum from the fields
$H_{\m\n}^{(00)}$, $M^{(00)}$, $N^{(00)}$ and $X^{5(00)}_{\m}$ is the
single massive mode described by \eq{ee5}, that is

\be
(\nabla^{2}_{\,x}-8)N^{(00)}\se 0\ .\la{snm}
\ee

To obtain this result, we first use the $AdS_{3}$ reparametrization
invariance \eq{diff} to fix the gauge

\be
\nabla^{\m}H_{\m\n}^{(00)}\se \ft{15}{4}\nabla_{\n}N^{(00)}\ .\la{gc}
\ee

We then obtain from \eq{ee2} the equation 

\be
(\nabla^{2}_{\,x}-3)M^{(00)}\se0\ .
\ee

In the gauge \eq{gc}, the residual reparametrizations obey

\bea
&&(\nabla^{2}_{\,x}-3)\nabla^{\m}\x_{\m}\se0\ ,
\nn\w2
&&(\nabla^{2}_{\,x}+2)\nabla_{\{\m}\x_{\n\}}
+\ft13\nabla_{\{\m}\nabla_{\n\}}\nabla^{\r}\x_{\r}\se0 \ .\la{rgt}
\eea

Using the trace
part $\nabla^{\m}\x_{\m}$, we set 

\be
M^{(00)}\se \ft92N^{(00)}\ .
\ee

Consequently \eq{ee1} takes the form

\be
(\nabla^{2}_{\,x} +2) H_{\m\n}^{(00)}\se 0\ .
\ee

The residual gauge transformations now obey

\be
(\nabla^{2}_{\,x} +2)\nabla_{(\m}\x_{\n)}\se0\ ,
\qquad \nabla^{\m}\x_{\m}\se0\ ,
\ee

and they allow us to impose the gauge 

\be
H_{\m\n}^{(00)}\se 0\ ,
\ee

with residual global reparametrizations generated by the Killing vectors
of $AdS_{3}$. Finally, the equation of motion \eq{sd1} determine
$X^{5(00)}_{\m}$ in terms of $N^{(00)}$ up to a homogeneous solution
described by a pure gauge degree of freedom which is removed completely
from the spectrum by the tensor gauge transformation \eq{diff} generated
by $\y^{5}$.

\item[$ii$)] From \eq{sd1} it follows that $X^{\ui\,(00)}_{\m}$ describes a
pure gauge field which is gauged away completely by the tensor gauge
transformations \eq{diff} generated by $\y^{\ui}$. 

\item[$iii$)] From \eq{sd3} and \eq{sce} it follows that $\f^{5r(00)}$ and
$X_{\m}^{r(00)}$ modulo the tensor gauge tarnsformations \eq{diff}
generated by $\y^{r}$ give rise to one scalar mode

\be
(\nabla^{2}_{\,x}-8)\f^{5r\,(00)}\se0 \ .\la{f5r}
\ee

\end{enumerate}


\item  {\bf The $(1\,,\,\pm1)$ Sector}


The two vector modes $K^{(1,\,\pm1)}_{\m}$ and $Z^{5(1,\,\pm1)}_{\m}$
transform under the the Yang-Mills symmetry \eq{ymt2}. The Lorentz gauge
condition does not follow from the Einstein equations of motion in this
sector due to the Killing equation \eq{zm}. The gauge invariant
equations of motion read 

\bea
&&
 \left(\nabla^{2}_{\,x} -2 \right)\,K_{\m}^{(1,\pm1)}-\nabla_{\m}
\nabla^{\n}K_{\n}^{(1,\pm1)} + 4e_{\m}{}^{\n\r}\del_{\n}Z^{5(1,\pm1)}_{\r}
\mp\,8 Z_{\m}^{5(1,\pm1)} \se 0\ , 
\la{kzfbo}\w2
&&e_{\m}{}^{\n\r}\del_{\n}Z^{5\,(1,\pm1)}_{\r}
\pm 2Z^{5\,(1,\pm1)}_{\m} +K_\m^{(1,\,\pm1)} \se0 \ .\la{kzfao} 
\eea

We use \eq{ymt2} to impose the Lorentz gauge

\be
\nabla^{\m}K^{(1,\,\pm1)}_{\m}\se 0\ ,\la{lgco}
\ee

with residual gauge symmetries generated by parameters obeying the
massless vector field equation

\be
(\nabla^{2}_{\,x}+2)\del_{\m}\L^{(1,\,\pm1)}\se0\ .\la{mlvf}
\ee

As a consequence the massless modes in this sector turn out to be pure
gauge, leaving only massive modes in the spectrum, as will be detailed
in section 5 using $AdS_{3}$ harmonic expansion. In the concluding
section we shall discuss further the gauge fields arising from the
isometries of $S^{3}$ and the generation of their masses.

\end{itemize}


\section{Fermionic Sector}


\subsection{ Harmonic Expansion on $S^{3}$ and Gauge Fixing }


We expand the fermionic fluctuations in harmonic spinors on $S^{3}$

\bea
\psi_{\m}(x,y\,)&=&\sum\psi_{\m}^{(\ell,\,\pm1/2)}(x)
Y^{(\ell,\,\pm1/2)}(y\,)
\ ,\nn\w2
\psi_{a}(x,y\,)&=& \sum\left[\l^{(\ell,\,\pm3/2)}(x)Y_{a}^{(\ell,\,\pm3/2)}(y\,)
+\left\{\z_{1}^{(\ell,\,\pm1/2)}(x)\nabla_{\{a\}}
+\z_{2}^{(\ell,\,\pm1/2)}(x)\c_{a}\right\}Y^{(\ell,\,\pm1/2)}(y\,)\right]
\ ,\nn\w2
\chi^{r}(x,y\,)&=&\sum\chi^{r\,(\ell,\,\pm1/2)}(x)
Y^{(\ell,\,\pm1/2)}(y\,)
\ ,\nn
\w3
\nabla_{\{a\}}&\equiv& \nabla_{a}-
\ft13\c_{a}\c^{b}\nabla_{b}\ .\la{ssp}
\eea

The spinor harmonics are Wigner functions in the coset $SO(4)/SO(3)$.
They carry a supressed row index transforming as a spinor of $SO(3)$.
The representation label $\ell$ is a positive half-integer subject to the
condition \eq{rest}. The spinor harmonics obey

\bea
\nabla_{y}^{2}Y^{(\ell,\,\pm1/2)}&=&\left[\ft32-(\ell+1)^{2}\right]
Y^{(\ell,\,\pm1/2)}\ ,\qquad \bs~Y^{(\ell,\,\pm1/2)}
\se\pm\,i\,(\ell+1)Y^{(\ell,\,\pm1/2)}
\ ,\nn\w2
\nabla_{y}^{2}Y^{(\ell,\,\pm3/2)}_{a}&=&\left[\ft52-(\ell+1)^{2}\right]
Y^{(\ell,\,\pm3/2)}_{a}\ ,\qquad \bs ~Y^{(\ell,\,\pm3/2)}_{a}
\se\pm\,i\,(\ell+1)Y^{(\ell,\,\pm3/2)}_{a}
\ ,\nn\w2
\c^{a}Y^{(\ell,\,\pm3/2)}_{a}&=&\nabla^{a}Y^{(\ell,\,\pm3/2)}_{a}\se0
\ .\la{sh}
\eea

The spinor harmonics $Y^{(\ell,\,\pm1/2)}$ form a complete set of
spinors on $S^{3}$. A complete set of vector-spinors on $S^{3}$ consists
of the transverse, $\c$-traceless vector-spinor modes
$Y_{a}^{(\ell,\,\pm3/2)}$, the longitudinal, $\c$-traceless modes
$\nabla_{\{a\}}Y^{(\ell,\,\pm1/2)}$ and $\c_{a}Y^{(\ell,\,\pm1/2),}$.
Note the zero modes

\be
\nabla_{\{a\}}Y^{(1/2,\,\pm1/2)}\se0\ .
\ee

The spinors $\psi_{M}$ and $\chi^{r}$ split into left- and right-handed
doublets of the unbroken $SO(4)_{R}\subset SO(5)$, labelled by the 
eigenvalue of $\C^{5}$ as follows 

\be
\C^{5}\psi_{M}^{(\a)}\se \a\psi_{M}^{(k)}\ ,\qquad
\C^{5}\chi_{r}^{(\a)}\se \a\chi_{r}^{(k)}\ ,\qquad k=\pm1\ .
\ee

Since there is no mixing between the left- and the right-handed
$SO(4)_{R}$ spinors in the equations of motion or transformation rules,
we will study the sectors with a fixed $\C^{5}$ eigenvalue separately
and supress the label $\a$ on the spinors. The zero modes
$Y^{(1/2,\,\a\,/2)}$ are identified with the $S^{3}$ component of the
Killing spinor \eq{ksst} in the sector with internal $SO(5)$ eigenvalue
$\a$. 

The linearized local supersymmetry
transformations read

\bea
\d \psi_{a}&=& \left(\nabla_{a}-\ft{i\,\a}{2}\c_{a}\right)\e
\ ,\nn\w2
\d\psi_{\m}&=&
\left(\nabla_{\m}+\ft{\a}{2}\c_{\m}\right)\e
\ ,\w2
\d \chi^{r}&=&0\ .\nn
\eea

By a field dependent supersymmetry transformation generated by the
harmonic non-zero modes and the zero mode $Y^{(1/2,\,-\a\,/2)}$ 
(which contribute to $\psi^{(\a)}_{a}$), we can remove all the 
terms in the expansion for the internal gravitino proportional
to $\c_{a}$ except for the zero-mode $\c_{a}Y^{1/2,\,\a\,/2)}$.
Thus we set to zero the following gauge modes:

\bea
\ell\geq\ft32&:&\quad \z_{2}^{\,(\ell,\,\pm1/2)}\se 0\ ,\w2
&&\quad \z_{2}^{\,(1/2,\,-\a\,/2)}\se0\ .\la{fgm}
\eea

This shows the admissibility of the gauge condition

\be
\psi_{a}\se \psi_{\{a\}}~+~\z^{\,(1/2,\,\a\,/2)}\c_{a}Y^{(1/2,\,\a\,/2)}
\ ,\qquad \c^{a}\psi_{\{a\}}\;\;\equiv\;\;0\ ,\la{psia}
\ee

where $\psi_{\{a\}}$ has an expansion in terms of $\c$-traceless
non-zero modes

\be
\psi_{\{a\}}\se\sum{}\left[\l^{(\ell,\,\pm3/2)}Y_{a}^{(\ell,\,\pm3/2)}
+\z^{\,(\ell,\,\pm1/2)}\nabla_{\{a\}}Y^{(\ell,\,\pm1/2)}\right]
\ ,\la{psiatl}
\ee

and we have set 

\bea
\ell\geq\ft32&:&\quad \z_{1}^{(\ell,\,\pm1/2)}\;\;\equiv\;\;\z^{(\ell,\,1/2)}
\ ,\nn\w2
&&\quad\z_{2}^{(1/2,\,\a\,/2)}\;\;\equiv\;\;\z^{(\ell,\,\a\,/2)}\ .
\eea

The gauge \eq{psia} completely fixes the local supersymmetries that are
spontaneously broken by the compactification, that is the symmetries
generated by the spinor harmonic non-zero modes $Y^{(\ell,\pm1/2)}$
($\ell\geq\ft32$) and the zero mode $Y^{(1/2,\,-\a\,/2)}$. The unfixed
zero modes $Y^{(1/2,\,\a\,/2)}$ generate the unbroken local $AdS_{3}$
supersymmetries acting on $\psi_{\m}^{(1/2,\,\a\,/2)}(x)$ as

\be
\d\psi_{\m}^{(1/2,\,\a\,/2)}(x)\se
\left(\nabla_{\m}+\ft{\a}{2}\c_{\m}\right)\e^{(1/2,\,\a\,/2)}(x)
\ .\la{lss}\w2
\ee

The gauge fixing of these local supersymmetries involves fixing of residual gauge
transformations that completely gauge away the massless gravitino from
the spectrum. After the gravitino has been gauged away there is a
remaining global supersymmetry that arranges the $AdS_{3}$ spectrum into
supermultiplets. The nature of the residual supersymmetry algebra and
the supermultiplet structure of the spectrum will be explained in
section 6.


{\small
\begin{table}[h]
\begin{center}
\begin{tabular}{|lll|c|c|c|}
\hline
&&&&&\\
&&Unbroken&&&\\
Sector:&&symmetry:&$\psi_{\m}$&$\psi_{a}$&$\chi^{r}$\\
&&&&&\\
\hline
&&&&&\\
$(\ell,\,\pm\ft32)\,,\ \ell\geq\ft32$&&-&&$\l$&\\
&&&&&\\
$(\ell,\,\pm\ft12)\,,\ \ell\geq\ft32$&&-&$\psi_{\m}$&$\z$&\\
&&&&&\\
$(\ft12,\,-\ft{\a}{2})$&&-&$\psi_{\m}$&&\\
&&&&&\\
$(\ft12,\,\ft{\a}{2})$&&$\e$&$\psi_{\m}$&$\z$&\\
&&&&&\\
$(\ell,\,\pm\ft12)\,,\ \ell\geq\ft12$&&-&&&$\chi^{r}$\\
&&&&&\\
\hline
\end{tabular}
\end{center}
\caption{{\small The $SO(4)$ representations arising in the harmonic expansion
of the fermionic fields in the gauge \eq{psia}. The $\C^{5}$ chirality
is denoted by $\a$. The fields in each row are described by a coupled
system of equations which must be diagonilized to find the spectrum. The
residual gauge symmetries are indicated by their parameters. }}

\la{tab2}
\end{table}
}


\subsection{ The Linearized Equations of Motion }


The linearized gravitino and fermion equations read

\bea
&&\c^{\m\n\r}\nabla_{\n}\psi_{\r}+i\,\c^{\m\n}\bs_{y}\,\psi_{\n}-
\c^{\m}\nabla^{a}\psi_{a}
+\a\,\c^{\m\n}\psi_{\n}-i\,\c^{\m\n}\nabla_{\n}\c^{a}\psi_{a}+
\c^{\m}\bs_{y}\c^{a}\psi_{a}\se0
\ ,\la{ge1}\w2
&&(i\bs_{x}+\bs_{y}\,)\psi^{a}+\c^{a}\left(
\c^{\m\n}\nabla_{\m}\psi_{\n}+\bs_{y}(i\c^{\m}\psi_{\m}+\c^{b}\psi_{b})-
\nabla^{b}\psi_{b}-i\bs_{x}\c^{b}\psi_{b}-i\a\c^{b}\psi_{b}\right)\nn\w1
&&\qquad\qquad -\nabla^{a}(i\c^{\m}\psi_{\m}+\c^{b}\psi_{b})
+i\a\psi^{a}\se0
\ ,\la{ge2}\w2
&&\left(\bs_{x}+i\,\bs_{y}-\a\,\right)\chi^{r}\se0\ ,\la{fe}
\eea

where we have used the decomposition of $\C^{M}$ under
$SO(1,5)\rightarrow SO(1,2)\times SO(3)$ given in Appendix A. Note that
the last two terms in the gravitino equations only contribute for the
zero-modes in \eq{psia}. Inserting the harmonic expansions for
$\psi_{\m}$ and $\chi^{r}$ given in \eq{ssp} and the expansion for
$\psi_{\{a\}}$ given in \eq{psiatl} into the linearized equations
\eq{ge1} - \eq{fe} and using 

\bea
\nabla^{a}\nabla_{\{a\}}Y^{(\ell,\,\pm1/2)}&=&
-\ft23\left[(\ell+1)^{2}-\left(\ft32\right)^{2}\right]Y^{(\ell,\,\pm1/2)}
\ ,\nn\w2
\bs_{y}\nabla_{\{a\}}Y^{(\ell,\,\pm1/2)}
&=& \ft{\pm\,i\,(\ell+1)}{3}\nabla_{a}Y^{(\ell,\,\pm1/2)}
-\ft49\left[(\ell+1)^{2}-\left(\ft32\right)^{2}\right]
\c_{a}Y^{(\ell,\,\pm1/2)}\ ,
\eea

we find the following set of irreducible equations (see Table
\ref{tab2} for the $SO(4)$ content of the fermionic modes):

\begin{itemize}


\item {\bf The $J=\ft32$ Sector}


$(\ell,\,\pm\ft12)$, $\ell\geq\ft32$ and $(\ft12,-\ft{\a}{2})$:
From \eq{ge1} and \eq{ge2} follows that 
$\psi_{\m}^{(\ell,\,\pm1/2)}$ gives rise to a massive spin $\ft32$ field
described by a transverse and $\c$-traceless vector-spinor

\bea
\vf_{\m}^{(\ell,\,\pm1/2)}&=&\psi_{\{\m\}}^{(\ell,\,\pm1/2)}
-\ft13{\ell+1\pm\ft{3\a}{2}\/\ell+1\mp\ft{\a}{2}}
\,\nabla_{\{\m\}}\z^{(\ell,\,\pm1/2)}\ ,\qquad \ell\geq\ft32
\nn\w2
\vf_{\m}^{(1/2,\,-\a\,/2)}&=&\psi_{\{\m\}}^{(1/2,\,-\a\,/2)}\ ,\la{mgr}
\eea

satisfying 

\bea
\left[\,\bs_{x}\pm(\ell+1)-\a\,\right]\vf_{\m}^{(\ell,\,\pm1/2)}&=&0\ ,\qquad
\ell\ge{\ft32}
\ ,\w2
\left(\,\bs_{x}-\ft52\,\a\,\right)\vf_{\m}^{(1/2,\,-\a\,/2)}&=&0\ .\la{f1}
\eea


\item {\bf The $J=\ft12$ Sector}


The diagonilized spin $\ft12$ eigenmodes are

\begin{enumerate}

\item[$i$)] $(\ell,\pm\ft32)$, $\ell\geq \ft32$: $\l^{(\ell,\,\pm3/2)}$
which from \eq{ge2} can be seen to obey

\be
\left[\bs_{x}\pm(\ell+1)+\a\right]\l^{(\ell,\,\pm3/2)}\se0\ .\la{lem}
\ee

\item[$ii$)] $(\ell,\pm\ft12)$, $\ell\geq \ft32$: 
$\z^{(\ell,\,\pm1/2)}$ which from \eq{ge1} and \eq{ge2} can be seen to obey

\be
\left[\bs_{x}\mp(\ell+1)-\a\right]\z^{(\ell,\,\pm1/2)}\se0\ .\la{zem}
\ee

\item[$iii$)] $(\ell,\pm\ft12)$, $\ell\geq \ft12$: 
$\chi^{r\,(\ell,\,\pm1/2)}$ which from \eq{fe} can be seen to obey

\be
\left[\bs_{x}\mp(\ell+1)-\a\right]\chi^{r\,(\ell,\,\pm1/2)}\se0\ .\la{chiem}
\ee

\end{enumerate}


\item {\bf The $(\ft12,\ft{\a}{2})$ Sector}


This sector contains the zero modes $\z^{(1/2,\,\a\,/2)}$ and the 
gravitino modes $\psi_{\m}^{(1/2,\,\a\,/2)}$ and from \eq{ge1} and
\eq{ge2} follows that they satisfy

\bea
&&\c^{\m\n\r}(\nabla_{\n}+\ft12\,\a\,\c_{\n})\psi_{\r}^{(1/2,\,\a\,/2)}
-3i\,\c^{\m\n}\nabla_{\n}\z^{(1/2,\,\a\,/2)}+3i\,\a\,\c^{\m}\z^{(1/2,\,\a\,/2)}
\se0\ ,\qquad\qquad\la{sso}\w2
&&\c^{\m\n}\nabla_{\m}\psi_{\n}^{(1/2,\,\a\,/2)}-\a\,\c^{\m}
\psi_{\m}^{(1/2,\,\a\,/2)}
-2i\,\bs_{x}\z^{(1/2,\,\a\,/2)}-i\,\a\,\z^{(1/2,\,\a\,/2)}\se0
\ ,\la{sst}
\eea

which are invariant under the local supersymmetry transformations
\eq{lss}. Combining \eq{sso} and \eq{sst} one finds that the zero modes obey 

\be
(\bs_{x}-\ft52\,\a\,)\z^{(1/2,\,\a\,/2)}\se0\ .\la{zmeq}
\ee

The rest of the sector consists of pure gauge degrees of freedom.

\end{itemize}


\section{Harmonic Expansion on AdS and the the Full Spectrum}


The matrix elements of the AdS group $SO(2,2)$ can be labelled by using
the maximal compact subgroup $O(2)_I\times O(2)_{II}$ as a basis. In
this basis, a unitary irreducible representation of $SO(2,2)$ is
labelled by $(E_0,s_{0})$ where $E_0$ is the lowest eigenvalue of the
energy operator $M_{03}$ that generates $O(2)_I$ and $s_{0}$ is the
of $M_{12}$ (helicity) of the state with lowest energy.

The other symmetries of the $AdS_3\times S^3$ compactified theory are:

\begin{itemize}

\item {the $SO(4)$ isometry group of the three sphere }

\item {the $SO(4)_R$ subgroup of the original $SO(5)$ $R$-symmetry group
	in $D=6$}

\item {the $SO(n)$ symmetry group inherited from $D=6$}

\end{itemize}

Thus, we shall denote a given state in the spectrum by

\be
D^{(\ell_1,\ell_2)}\ (E_0,s_{0})\ (R\times S\,) \la{drs}
\ee

where $(\ell_1,\ell_2)$ label the $S^3$ isometry group $SO(4)$;
$(E_0,s_{0})$ label the representation of the AdS group $SO(2,2)$; 
$R$ denotes the representation of $SO(4)_R$ and $S$ denotes a
representation of $SO(n)$.

The internal $SO(4)$ is isomorphic to $SU(2)_L \times SU(2)_R$ with isospins 
$(j, {\bar j})$ related to the $SO(4)$ highest weight labels $(\ell_1,\ell_2)$ as

\be
j \se \ft12 (\ell_1+\ell_2)\ ,\qquad {\bar j} \se \ft12 (\ell_1-\ell_2)\ .
\la{jjb}
\ee

Similarly $SO(2,2) \approx SU(1,1)_L\times SU(1,1)_R$ with conformal
weights $(h,{\bar h})$ given by

\be
h\se \ft12 (E_0+s_{0})\ ,\qquad {\bar h} \se \ft12 (E_0-s_{0})\ .
\la{hhb}
\ee

To determine the $SO(2,2)$ content of the spectrum, we shall follow the
techniqe used in \cite{es1,es2} which is based on the analytic
continuation of $AdS_3$ to a three-sphere $S^3_{E}$, and consequently
the group $SO(2,2)$ to $SO(4)_E$. The Casimir eigenvalues for an
$SO(2,2)$ representation $D(E_0,s_{0})$ and an $SO(4)_E$ representation
with heighest weight $(n_1,n_2)$ are 

\bea
SO(2,2)\ : \qquad C_2 = E_0(E_0-2) +s_{0}^2 
\nn\w2
SO(4)_E\ : \qquad C_2= n_1(n_1+2) + n_2^2\ .\la{cast}
\eea

This suggests that in continuing from $S^3_{E}$ back to $AdS_3$ one
makes the identifications

\be
n_1=-E_0 \ , \quad\quad n_2=s_{0}\ ,
\ee
 
where, without loss of generality, we have chosen the plus sign in the
last relation. There are a number of subtleties involved in the
analytical continuation of $AdS_3$ to $S^3$. One particular consequence
is that one performs the change of coordinates $(x^1,x^2)\rightarrow
(ix^1,ix^2)$, and therefore the metric on $S^3_{E}$ has signature
$(---)$ and the replacement

\be
{\bar g}_{\m\n}\ \rightarrow \ -\bar{g}^{E}_{\m\n}\ ,\qquad \nabla^{2}_{\,x}
\rightarrow -\nabla^{2}_{_{E}}
\ee

needs to be made. A number of other issues involved here are direct analogs
of those discussed in detail in \cite{es1,es2} for the case of $AdS_4$
continued to $S^4$. We shall not repeat those discussions here but we
shall outline the salient features of the harmonic expansions on $AdS_3$
which lead to the determination of the complete spectrum. 

We need to expand scalars, vectors and symmetric traceless tensors on
$AdS_3$ in terms of suitable harmonic functions. Upon analytical
continuation to $S^3_{E}$ these expansions become exactly like those listed
in \eq{n} where the the AdS vector indices $(\m,\n,...)$ are to be treated
like the $S^3$ vector indices $(a,b,..)$. In other words, one makes the
replacements

\be
Y_{( s)}^{(\ell_1\ell_2)}\quad \rightarrow \quad 
Y_{( s_{0})}^{(n_1 n_2)}\ ,\qquad n_1\, \geq\, |n_2|\ .\la{repl}
\ee

Formulae such as \eq{yn1} and \eq{sh} can of course be used with
suitable renaming of the labels. Indeed, the fact that the harmonic
expansions are being performed in $S^3_{E}\times S^3$ simplifies the
calculations considerably since the same formula can be used for both of
the three spheres.

Next, we substitute the $AdS_3 \rightarrow S^3_{E}$ harmonic expansions
in all the diaonalized field equations listed in sections 3.3 and 4.2.
Let us label the $SO(4)_E$ representations involved by $(n n_2)$. In
fact, only the $0, \pm 1, \pm 2$ values of $n_2$ occur in the harmonic
expansions. We then determine the critical values of $n$ for which the
linearized wave operators vanish. Continuing back to $AdS_3$, we make
the identification $n=-E_0$, and interpret the value of $E_0\geq 1$ as
the lowest weight of the $AdS_3$ representation described by the field.
The significance of the value $E_{0}=1$ is due to the fact that it is
the fixed point of the interchange $E_{0}\rightarrow 2-E_{0}$ which
leaves the wave equations invariant. Therefore the solutions of the
field equations for $E_{0}<1$ can be obtained from those with $E_{0}>1$
by this interchange and the two solutions together characterize the
representation that we label with $(E_{0},s_{0})$. For first order
equations, such as Dirac equations and the Hodge duality conditions,
there is an additional flip of helicity $s_{0}\rightarrow -s_{0}$ to be
taken into account as well. This procedure will become transparent as we
present its application to various sectors.

\begin{itemize}


\item {\bf The $J=2$ Sector}


From \eq{ste} and \eq{mstm}, using the formula \eq{dely} and the
replacement \eq{repl}, we obtain

\be
\left[ n(n+2) +1-(\ell+1)^2 \right] S^{(\ell \,0)(n,\,\pm2)} =0\ ,
\qquad n\geq 2\ ,\quad \ell \geq 1\ ,
\ee

where $S^{(\ell \,0)(n\pm2)}$ is the coefficient of the $S^{3}_{E}$
harmonic function $Y^{(n,\pm2)}(x_{E})$. The critical values of $n$ are
$n_{\pm}=-1\pm(\ell+1)$. Continuing back to $AdS_3$ by letting $n
\rightarrow -E_0$, we find that $E_0=\ell+2$ is the lowest energy of the
$AdS_3$ representation $D(E_0=\ell+2,s_{0}=\pm2)$. Using the notation
\eq{drs}, we then fully characterize the $J=2$ tower of states as

\be
D^{(\ell+1,\,0)}\,(\ell+3,\,\pm 2)\,(0,0)\ ,  \qquad \ell \geq 0 \ .\la{j2}
\ee

Note that we have shifted $\ell$ by one so that its minimum value is
zero. Subsequently we shall do so when necessary, so that $\ell$ will
serve as the level number. It is also worth noting that the lowest
member of the spin 2 tower exhibited in \eq{j2} is a massive vector of
$SO(4)$ internal symmetry group. The massless graviton does not arise in
the physical spectrum, as expected.


\item {\bf The $J=\ft32$ Sector}


Starting from \eq{sh} and making the replacement \eq{repl} we obtain

\be
\bs_{E}Y_{\m}^{(n_{1},\,\pm3/2)}(x_{E})\se \pm\,i\,(n_{1}+1)
Y_{\m}^{(n_{1},\,\pm3/2)}(x_{E})\ ,\qquad n_{1}\ge3/2\ .\la{adsth}
\ee

Continuing the gravitino equations \eq{mgr} from $AdS_{3}$ to
$S^{3}_{E}$ using technique spelled out above, and the replacement 

\be
\c^{\m}_{E}\rightarrow i\c^{\m},
\ee

we find critical values of the energy and the helicity implying 
that the vector-spinor $\vf_{\m}$
contains the representations

\bea
D^{(\ell+3/2,\,\pm1/2)}(\ell+\ft52,\pm\ft32)(2_{\pm},0)\ ,\qquad
\ell\geq0
\ ,\nn\w3
D^{(\ell+1/2,\,\pm1/2)}(\ell+\ft72,\pm\ft32)(2_{\mp},0)\ ,\qquad
\ell\geq0\ .\la{j32}
\eea

The correlation between the $AdS_{3}$ helicity $s_{0}$ and the $SO(4)$
helicity $l_{2}$ is a consequence of chosing the positive energy root.


\item {\bf The $J=1$ Sector}


Starting from the last equation in \eq{yn1} and making the replacement
\eq{repl} we obtain

\be
e^{^{_{E}}\ }_{\m}{}^{\n\r}\del_{\n}Y_{\r}^{(n_{1},\,\pm1)}(x_{E})\se
\pm(n_{1}+1)Y_{\m}^{(n_{1},\,\pm1)}(x_{E})\ ,\qquad n_{1}\ge1\ .
\ee

Continuing back to $AdS_{3}$ and using the replacement 

\be
e^{^{_{E}}}_{\m\n\r}\rightarrow -e_{\m\n\r}\ ,
\ee

we find that the resulting mode functions $Y_{\m}^{(E_{0},s_{0})}(x)$
with energy $E_{0}$ and helicity $s_{0}$ obey the relation

\be
\e_{\m}{}^{\n\r}\del_{\n}Y_{\r}^{(E_{0},\,\pm1)}\se
\mp(E_{0}-1)Y_{\m}^{(E_{0},\,\pm1)}\ ,\qquad E_{0}\ge1\ .\la{curld}
\ee

From the equations \eq{kzfb}, \eq{kzfa} \eq{lgc} for
$K_\m^{(\ell,\pm1)}$ and $Z_{\m}^{5(\ell,\,\pm1)}$ ($\ell\ge2$) and from the
equations \eq{kzfbo}, \eq{kzfao} and \eq{lgco} for $K_\m^{(1,\pm1)}$ and
$Z_{\m}^{5(1,\,\pm1)}$, and using \eq{curld} and the last equation in
\eq{yn1}, we find the critical $E_{0}$ values which correspond to the representations 

\bea
&&
D^{(\ell+2,\,\pm1)}\,(\ell+2,\,\pm1)\,(0,0)\ ,  \qquad \ell \geq 0\ ,
\la{j1c}\w2
&&
D^{(\ell+1,\,\pm 1)}\,(\ell+3,\,\mp 1)\,(0,0)\ ,
\qquad \ell \geq 0 
\ ,\la{j1a}\w2
&&
D^{(\ell+1,\,\pm1)}\,(\ell+5,\,\pm1)\,(0,0)\ .  \qquad \ell \geq 0 
\ ,\la{j1b}
\eea

For each label $(\ell,\,\pm1)$ ($\ell\geq 2$) the system thus yields
three degrees of freedom. This can be understood as follows. A single
vector field obeying a {\it second order} Proca equation together with a
Lorentz gauge condition gives rise to {\it two} degrees of freedom; one
with helicity $s_{0}=1$ and one with helicity $s_{0}=-1$. On the other
hand, a single vector $V_{\m}$ obeying the {\it first order} equation
$e^{\m\n\r}\del_{\n}V_{\r}+M\,V^{\m}=0$, where $M$ is a mass parameter,
gives rise to {\it one} degree of freedom with energy $E_{0}=1+|M|$ and
helicity $s_{0}=M/|M|$. 

In our case there is an additional contribution to the the second order
equation \eq{kzfb} in the form of a first order derivative term, which
splits the energy levels of the two critical solutions. The direction of
the energy shift is determined by the sign of the internal $SO(4)$
helicity label $\ell_{2}$. Also the sign of the mass parameter $M$ of
the first order equation introduced in the previous paragraph is
determined by the sign of $\ell_{2}$, as can be seen from \eq{kzfa}. As
a result, in all the three towers of representations in \eq{j1c},
\eq{j1a} and \eq{j1b}, the helicity $s_{0}$ of a positive energy
solution is correlated with the $SO(4)$ helicity label $\ell_{2}$.

For $(1,\pm1)$, $K_\m^{(1,\pm1)}$ and $Z_{\m}^{5(1,\,\pm1)}$ gives rise to
four critical solutions; $(E_{0}=1,s_{0}=\ell_{2})$,
$(E_{0}=1,s_{0}=-\ell_{2})$ $(E_{0}=3,s_{0}=-\ell_{2})$ and
$(E_{0}=5,s_{0}=\ell_{2})$. The equation \eq{curld} implies that the
$E_{0}=1$ modes are total derivatives, which can be gauged away using
the residual gauge parameter that satisfy \eq{mlvf}. As a result the
tower \eq{j1c} starts with a $(2,\pm)$ representation of $SO(4)$, and
the two physical Yang-Mills states in the adjoint representations
$(1,\pm 1)$ of $SU(2)_{L,R}$ sit at the bottom of the two towers
\eq{j1a} and \eq{j1b}.\footnote{
In $3D$, the Yang-Mills system $\nabla_{\n}F^{\n\m}+Me^{\m\n\r}F_{\n\r}=0$, where 
$F_{\m\n}=\del_{\m}A_{\n}-\del_{\n}A_{\m}$ and $M\neq 0$ is a topological mass parameter,
has three critical solutions; $(E_{0},s_{0})=(1,\pm1)$ and 
$(E_{0},s_{0})=(1+2|M|,-M/|M|)$. The $E_{0}=1$ modes are pure gauge
and the system thus has one physical state with topological mass. The compactified
theory happens to fix the value of $M$ in terms of the cosmological
constant.}\la{pr1}

From the first order equation \eq{zui} we find that the
representation content of $Z_\m^{\ui\,(\ell,\,\pm 1)}$  is

\be
D^{(\ell+1,\,\pm 1)}\,(\ell+3,\,\pm 1)\,(4,0)\ ,  \qquad \ell \geq 0 \ .
\la{j1d}
\ee

Finally, the representation content of $Z_\m^{r\,(\ell,\,\pm 1)}$ is found
from the first order equation \eq{zr} to be 

\be
D^{(\ell+1,\,\pm 1)}\,(\ell+3,\,\mp 1)\,(0,n)\ , \qquad \ell \geq 0 \ .
\la{j1dn}
\ee


\item {\bf The $J=\ft12$ Sector}


From \eq{lem} and the relation

\be
\bs_{x}Y^{(E_{0},\,\pm1/2)}(x_{E})\se 
\mp(E_{0}-1)Y^{(E_{0},\,\pm1/2)}(x_{E})
\ ,
\ee

we find that the spinor $\l$ contains the representations

\bea
&&D^{(\ell+3/2,\,\pm3/2)}(\ell+\ft92,\pm\ft12)(2_{\pm},0)\ ,\qquad
\ell\geq0
\ ,\nn\w3
&&D^{(\ell+3/2,\,\pm3/2)}(\ell+\ft52,\pm\ft12)(2_{\mp},0)\ ,\qquad
\ell\geq0\ .\la{j12a}
\eea

From \eq{zem} and \eq{zmeq} we find that the spinor $\z$
contains the representations

\bea
&&D^{(\ell+1/2,\,\pm1/2)}(\ell+\ft72,\mp\ft12)(2_{\pm},0)\ ,\qquad
\ell\geq0
\ ,\nn\w3
&&D^{(\ell+3/2,\,\pm1/2)}(\ell+\ft52,\mp\ft12)(2_{\mp},0)\ ,\qquad
\ell\geq0\ .\la{j12b}
\eea

From \eq{chiem} we find that the spinor $\chi^{r}$
contains the representations

\bea
&&D^{(\ell+1/2,\,\pm1/2)}(\ell+\ft72,\mp\ft12)(2_{\mp},n)\ ,\qquad
\ell\geq0
\ ,\w3
&&D^{(\ell+1/2,\,\pm1/2)}(\ell+\ft32,\mp\ft12)(2_{\pm},n)\ ,\qquad
\ell\geq0\ .\la{soltwo}
\eea

Note that for $\ell=0$ \eq{soltwo} actually has a solution with positive
energy $E_{0}=\ft12$, but as explained earlier this solution only
characterize the field configurations obtained by the interchange
$E_{0}\rightarrow 2-E_{0}$, $s_{0}\rightarrow -s_{0}$, and do not
represent new states in the spectrum.


\item {\bf The $J=0$ Sector}


From \eq{ynpm}, \eq{wat} and \eq{snm} we find that $N_{-}^{(\ell\,0)}$ 
($\ell\geq1$) and
$N^{(00)}$ together form the representation tower

\be
D^{(\ell,\,0)}\,(\ell+4,\,0)\,(0,0)\ .  \qquad \ell \geq 0 \la{j0a}
\ee

Similarly, from \eq{ynpm} it follows that the field $N_{+}^{(\ell\,0)}$
($\ell\geq2$) describes the representations

\be
D^{(\ell+2,\,0)}\,(\ell+2,\,0)\,(0,0)\ ,  \qquad \ell \geq 0 \ .\la{j0c}
\ee

It follows from \eq{lpt} that $L^{(\ell,\,\pm 2)}$ has the
representation content

\be
D^{(\ell+2,\,\pm2)}\,(\ell+4,\,0)\,(0,0)\ ,  \qquad \ell \geq 0 \ .\la{j0d}
\ee

Next, from \eq{fsm} we find the representation content of the field
$U^{\ui\,(\ell\,0)}$ ($\ell\geq1$) to be

\be
D^{(\ell+1,\,0)}\,(\ell+3,\,0)\,(4,0)\ ,  \qquad \ell \geq 0 \ .\la{j0e}
\ee

From \eq{ynpm2} and \eq{f5r} follows the two towers of representations
obtained from $U^{r(\ell0)}$ ($\ell\geq1$) and $\f^{5r(\ell0)}$
($\ell\geq0$):

\bea
&&D^{(\ell,\,0)}\,(\ell+4,\,0)\,(0,n)\ ,  \qquad \ell \geq 0 \ ,\la{j0f}
\w2
&&D^{(\ell+1,\,0)}\,(\ell+1,\,0)\,(0,n)\ , \qquad \ell \geq 0 \ .\la{j0g}
\eea

Finally, from \eq{fosm} we find that
$\phi^{\ui\,r\,(\ell\,0)}$ has the representation content

\be
D^{(\ell,\,0)}\,(\ell+2,\,0)\,(4,n)\ ,  \qquad \ell \geq 0 \la{j0h}
\ee

\end{itemize}


\section{The Supermultiplet Structure of the Spectrum}


Applying harmonic analysis to the Killing equation \eq{ksst} we find
that the Killing spinors, associated with the supercharges $Q_+$ and
$\overline{Q}_-$, contain the representations

\bea
&& 
Q_+\ : \quad D^{(1/2,\,1/2)}(-\ft12,\,-\ft12)(2_{+},\,0)\ , \nn
\w2
&& \overline{Q}_-\ : \quad D^{(1/2,\,-1/2)}(-\ft12,\,\ft12)(2_{-},\,0)\ .
\la{q12}
\eea

These supercharges obey

\bea
{}[E,\,Q_{+}\,]&=&-\ft12 Q_{+}\ ,\qquad [E,\,\overline{Q}_{-}\,]\se-
\ft12\overline{Q}_{-}
\ , \nn\w2
{}[J,\,Q_{+}\,]&=&-\ft12 Q_{+}\ ,\qquad [J,\,\overline{Q}_{-}\,]\se
\ft12\overline{Q}_{-}
\ , \la{scharges}\w2
\C^{5}Q_{+}&=&Q_{+}\ ,\qquad\qquad
\overline{Q}_{-}\C^{5}\se-\overline{Q}_{-}\ ,\nn
\eea

where $E$ and $J$ are the hermitian energy and the helicity operators of
$SO(2,2)$. Taking the hermitian conjugates of these equations we find
that the supercharges $\overline{Q}_{+}$ and $Q_{-}$ carry the
representations

\bea
&& \overline{Q}_+ \ :\quad D^{(1/2,\,1/2)}(\ft12,\,\ft12)(2_{+},\,0)\ ,
\nn\w2
&& Q_- \ :\quad D^{(1/2,\,-1/2)}(\ft12,\,-\ft12)(2_{-},\,0)\ ,\la{erq}
\eea

and obey the commutation rules

\bea
{}[E,\,\overline{Q}_{+}\,]&=&\ft12 \overline{Q}_{+}
\ ,\qquad [E,\,Q_{-}\,]\se\ft12 Q_{-}
\ , \nn \w2
{}[J,\,\overline{Q}_{+}\,]&=&\ft12 \overline{Q}_{+}
\ ,\qquad [J,\,Q_{-}\,]\se-\ft12 Q_{-}
\ ,\la{schargeshc} \w2
\overline{Q}_{+}\C^{5}&=&Q_{+}\ ,\qquad\qquad 
\C^{5}Q_{-}\se-\overline{Q}_{-}\ . \nn
\eea

The full algebra satisfied by the supercharges $Q_{\pm}$ and
$\overline{Q}_{\pm}$ is

\be
SU(1,1|2)_{L}\oplus SU(1,1|2)_{R}\ ,
\ee

\la{pref} where the bosonic subalgebras are $\left(SU(1,1)\oplus
SU(2)\right)_{L,R}$ (see the Appendix where the algebra is given). The
$SU(2)_{L}\times SU(2)_{R}$ is the isometry group of the internal
$S^{3}$. Representing the superalgebra in this way requires that the
supercharges $Q_L$ are Dirac and that they carry the doublet indices of
$SU(1,1)_L\times SU(2)_L$. Hence there are four complex supercharges in
the left-handed sector. One can go over to a real basis by introducing a
doublet index which transforms under a global $SU(2)_+$ group that
commutes with the superalgebra. A similar step in the right-handed
sector yields an $SU(2)_-$. Together $SU(2)_+ \times SU(2)_-$ can be
interpreted as a the global $SO(4)_R$ inherited from the $R$ symmetry
group $SO(5)$ in $D=6$. We emphasize that the generators of $SO(4)_R$ do
not arise on the right hand side of the superalgebra.

To find the supermultiplets structure of the spectrum we begin by acting
with the {\it helicity lowering} supercharges $Q_{\pm}$ on the tower of
highest helicity representations in the physical spectrum, that is the
representations

\be
D^{(\ell+1,\,0)}\,(\ell+3, 2)\,(0,0)\la{j2p}
\ee

found in \eq{j2}\footnote{ Alternatively, the same supermultiplet
structure can be found by first identifying the ground state of minimal
energy $E_{0}$ and then act on the ground state with the {\it energy
raising} supercharges $\overline{Q}_{+}$ and $Q_{-}$ given in
\eq{erq}.}. Using the $SO(4)$ tensor product rule

\be
(\ell_{1},\,\ell_{2})\otimes (\ft12,\,\pm\,\ft12)\se
(\ell_{1}+\ft12,\,\ell_{2}\pm\ft12)\oplus
(\ell_{1}-\ft12,\,\ell_{2}\mp\ft12)\ ,
\ee

and discarding representations not in the fully antisymmetrized product,
we can combine the representations \eq{j2p}, \eq{j32}, \eq{j1c}, 
\eq{j1b}, \eq{j1d}, \eq{j12a} and \eq{j0d} into the tower of spin $2$
supermultiplets given in Table \ref{htt} and the supersymmetry
transformation rules are shown in Figure \ref{httf}. This tower contains
all $|s_{0}|>1$ representations and all the $s_{0}=1$ representations
except the $SO(4)\times SO(n)$ singlets in \eq{j1a}, that is the
representations 

\be
D^{(\ell+1,\,-1)}(\ell+3,\,1)(0,0)\ ,\la{tt}
\ee 

and the $SO(4)\times SO(n)$ $(0,n)$-plets in \eq{j1dn},
that is the representations

\be
D^{(\ell,\,\pm1)}(\ell+3,\mp1)(0,n)\ .\la{ttt}
\ee

Acting with the helicity lowering supercharges $Q_{\pm}$ on \eq{tt}
we find that \eq{ttt} together with the representations 
\eq{j12b}, \eq{j0c}, \eq{j0e} and \eq{j0a} fit into the two tower of
supermultiplets given in Table \ref{hot}, with 
supersymmetry transformation rules in Figure \ref{hotf}. Similarly \eq{ttt}, 
\eq{soltwo}, \eq{j0g}, \eq{j0h} and \eq{j0f} constitute the tower 
\ref{hotn}, with supersymmetry transformation rules in Figure \ref{hotnf}. 

Note that by repeated action of $Q_{\pm}$ starting from the helicity
$+2$ graviton state one can only reach the helicities $s_{0} \geq 0$.
The reason for this is that the tower of ``massive" spin $2$ multiplets
originates from the {\it shortened}, massless $D=6$ gauge multiplets of
the infinitely many gauge symmetries that are spontaneously broken by
the $AdS_{3}\times S^3$ vacuum \cite{dnp}. Technically speaking, the
momenta of the internal $S^3$ that are picked up by the $SO(4)$
generators in the right hand side of the superalgebra implies that the
set of fermionic creation operators formed out of the supercharges
$Q_{\pm}$ can be applied at most twice to physical states, as can be
seen from Figures \ref{httf}, \ref{hotf} and \ref{hotnf}. 

Thus, the two spin $1$ towers consist of self-conjugate spin $1$
supermultiplets. In the case of the spin $2$ tower, the full multiplet
structure is obtained by adding the conjugate tower of multiplets with
the replacements $\ell_{2}\rightarrow -\ell_{2}$, $s_{0}\rightarrow
-s_{0}$, $2_{\pm}\rightarrow 2_{\mp}$ made. This conjugate tower of
supermultiplets can be obtained by repeated action of
$\overline{Q}_{\pm}$ starting from the helicity $-2$ graviton state.


The spin $2$ tower of physical states shown in Figure \ref{httf} are
labelled by a level number $\ell$ which starts from zero. If one
extrapolates to $\ell=-1$, one finds the nonpropagating supergravity
multiplet in the upper left diagonal and its conjugate image consisting
of a graviton, gravitini in the $(2_{L},2_{+})$ and $(2_{R},2_{-})$
representations of $SO(4)\times SO(4)_{R}$ and $SO(4)$ vector fields.

For $\ell=0$, the states on the lower right diagonal are absent because
of the group theoretical restriction $\ell_1 \geq |\ell_2|$, on the
$SO(4)$ representation labels $(\ell_1,\ell_2)$. Thus one is left with a
spin $2$ multiplet with 48 Bose and 48 Fermi degrees of freedom and
lowest spin $\ft12$.

For $\ell \geq 1$ the supermultiplet structure is generic, and has a
total of $16(\ell+1)(\ell+3)$ Bose and $16(\ell+1)(\ell+3)$ Fermi
degrees of freedom. Note however, that the $SO(4) \approx SU(2)_L \times
SU(2)_R$ Yang-Mills fields originating from the isometries of the
internal $S^3$ reside at level $\ell=1$ multiplet with $128$ Bose and
$128$ Fermi degrees of freedom. Furthermore, the helicity $+1$
Yang-Mills states are in the adjoint representation of $SU(2)_L$ while
the helicity $-1$ Yang-Mills states carry the adjoint representation of
$SU(2)_R$. The complementary helicity states are found in the spin $1$
tower of $SO(n)$ singlets.

The $D=6$ origin of various members of the spin $2$ tower is as follows.
The spin $2$ and $\ft32$ members, of course, come from the metric
$g_{\m\n}$ and the gravitino $\psi_\m$. The singlet spin $1$ states
involve a mixing of the Kaluza-Klein vectors $g_{\m a}$ and the vectors
$B^5_{\m a}$ originating from the self-dual tensors. The spin $1$ states
in the $4$ of $SO(4)_{R}$ come from the vectors $B^{\ui}_{\m a}$. The
spin $\ft12$ states come from the internal traceless and transverse
components of the internal gravitino field $\psi_a$. Finally, the
scalars originate from the internal metric $g_{ab}$. 


The spin $1$ tower of $SO(n)$ singlet states in figure \ref{hotf} starts
at level $\ell=0$. The level $0$ multiplet which has $32$ Bose and $32$
Fermi degrees of freedom is special since it contains the the Yang-Mills
states in the adjoint of $SU(2)_{R}$ with helicity $+1$ and in the
adjoint of $SU(2)_{L}$ with helicity $-1$ that are complementary to
those in the spin $2$ tower.

At level $l\geq1$ the supermultiplet is generic and it has a total of 
$8(\ell+2)^{2}$ Bose and $8(\ell+2)^{2}$ Fermi degrees of freedom.

The $D=6$ origin of the states is as follows. The spin $1$ states come
from the mixture between the Kaluza-Klein vector $g_{\m a}$ and the
selfdual tensor component $B^{5}_{\m a}$. The spin $\ft12$ states
originate from the longitudinal part of the internal graviton $\psi_a$.
The scalar $SO(4)_{R}$ $4$-plet come from the internal components
$B^{\ui}_{ab}$ of the selfdual tensor and the two scalar $SO(4)_{R}$
singlets come from the mixture between $g^{\m}{}_{\m}$, $g^{a}{}_{a}$
and $B^{5}_{ab}$. 

The spin $\ft12$ multiplet residing at the left diamond for $\ell=-1$
does {\it not} occur in the spectrum; the scalar $SO(4)_{R}$ singlet in
the $(10)$ of $SO(4)$ are the scalar states gauged away by the residual
Stueckelberg shift transformations; the scalar $SO(4)_{R}$ $4$-plet in
the singlet representation of $SO(4)$ never occured in the harmonic
expansion of $B^{\ui}_{ab}$ (see \eq{n}); and finally the $SO(4)_{R}$
doublet of fermions in the $(\ft12,\pm\ft12)$ of $SO(4)$ is put equal to
zero in the gauge \eq{psia} for the local supersymmetry. 


The spin $1$ tower of $SO(n)$ singlet states in figure \ref{hotf} is a
replica of the singlet spin $1$ tower except that this tower starts at
level $\ell=-1$. At this level the spin $\ft12$ multiplet residing in
the left diamond survives. This multiplet contains $8$ Bose and $8$
Fermi degrees of freedom. 

The $D=6$ origin of the states is as follows. The spin
$1$ states come from the anti-selfdual tensor $B^{r}_{\m a}$. The spin
$\ft12$ states come from the $SO(n)$ $n$-plet $\chi^{r}$ of chiral
fermions. The scalars in the $(4,n)$ of $SO(4)_R\times SO(n)$ come from
the coset scalars $\f^{\ui r}$. The remaining scalars come from the
mixture between $\f^{5r}$ and the internal components $B^{r}_{ab}$ of
the anti-selfdual tensor.


\section{Discussion}


The complete spectrum of $AdS_3 \times S^3$ compactified $D=6,\,N=4b$
supergravity which we have determined in this paper shares many
similarities to other known AdS compactified supergravity theories in
the literature. A new feature is that the spectrum contains more than
one tower of supermultiplets. This is not too surprising since the
underlying supersymmetry contains only $16$ real supercharges, which is half
of the maximal number of supersymmetries that arise in other $AdS$
compactified supergravity theories studied in the past. 

As expected, the massless supergravity sector of the compactified theory
is non-propagating. Other gauge modes one normally expects in $AdS$
compactifications of supergravity theories are the
singletons/doubletons. For example, in $AdS_4 \times S^7$
compactification of $D=11$ supergravity singletons arise in the
$\ell=-1$ level of the massive spin $2$ tower \cite{es1,es2}. In our
case, the supergroup $SU(1,1|2)_L \times SU(1,1|2)_R$ does not have
singletons \cite{g2}. However, the massless representations of the group
can be interpreted as doubletons. The self-conjugate such multiplet can
be found in \cite{g}. The representation with lowest energy in this
multiplet is $D^{(0,0)}(1,0)(0,0)$. This representation does not seem to
arise here.

The Yang-Mills sector of the theory is rather interesting. They gauge
the $SO(4) \approx SU(2)_L \times SU(2)_R$ isometry group of the full
$SO(4)\times SO(4)_R$ R-symmetry group of the superalgebra (see
discussion on p. \pageref{pref}). The equations of motion \eq{kzfbo} and
\eq{kzfao} prior to gauge fixing can be written as

\bea
&&\nabla_{\n}F^{\n\m}+2e^{\m\n\r}G_{\n\r}\se0\ ,\la{absa}\w2
&&e^{\m\n\r}G_{\n\r}+4(A^{\m}+B^{\m})\se0\ ,\la{absb}
\eea

where we have used the field strengths

\be
F_{\m\n}\se \del_{\m}A_{\n}-\del_{\n}A_{\m}\ ,\qquad
G_{\m\n}\se\del_{\m}B_{\n}-\del_{\n}B_{\m}\ ,\la{fst}
\ee

where we have set $A_{\m}\equiv K_{\m}^{(1,1)}$,
$B_{\m}\equiv 2\,Z_{\m}^{5\,(1,1)}$ and we have suppressed the adjoint
$SU(2)_{L}$ indices. The gauge transformations \eq{ymt2}
leaving \eq{absa} and \eq{absb} invariant are 

\be
\d A_{\m}\se \del_{\m}\L\ ,\qquad \d B_{\m}\se -\del_{\m}\L\ .\la{ymt3}
\ee

The equations \eq{absa} and \eq{absb} can be
derived from the Lagrangian

\be
e^{-1}{\cal L}\se -\ft14 F_{\m\n}F^{\m\n}-e^{\m\n\r}G_{\m\n}B_{\r}
-4(A_{\m}+B_{\m})(A^{\m}+B^{\m})\ .\la{l1}
\ee

The non-abelian version is obtained using the field strengths
$F=dA+A\wedge A$ and $G=dB-B\wedge B$ (the sign difference leads to that
the mass term remains invariant under the non-abelian gauge
transformations). The essential property of this action is that a single
gauge symmetry is associated with {\it two} gauge fields (note that if
one uses the gauge symmetry \eq{ymt3} to impose the Lorentz gauge
condition on $A_{\m}$ or $B_{\m}$, say $\nabla^{\m}A_{\m}=0$, then the
condition $\nabla^{\m}B_{\m}=0$ follows from the first order equation
\eq{absb}). Consequently this system describes only two degrees of
freedom on-shell (one from each vector), as shown in detail in the
treatment of the $J=1$ sector section 5.

Note that one can take the curl of the first order equation \eq{absb}
and diagonilize the resulting system of coupled second order equations
by going over to the field strengths $F_{1}=F+2G$ and
$F_{2}=\ft13(F-G)$. The resulting equations can be derived from an
action which is the sum of two actions each one containing a
Chern-Simons term {\it and} a kinetic term (causing propagation in the bulk):

\be
e^{-1}{\cal L}'\se -\ft14
F_{1\,\m\n}F^{\m\n}_{1}+\e^{\m\n\r}F_{1\,\m\n}A_{1\,\r}- \ft14
F_{2\,\m\n}F^{\m\n}_{2}-\ft12 \e^{\m\n\r}F_{2\,\m\n}A_{2\,\r}\ .\la{l2}
\ee

This action yields the same on-shell content as \eq{l1} {\it since}
the action \eq{l2} has {\it two } independent gauge
symmetries (see footnote on page \pageref{pr1})

\be
\d A_{1\,\m}\se \del_{\m}\L_{1}\ ,\qquad \d A_{2\,\m}\se
\del_{\m}\L_{2}\ .
\ee

This would however imply the enlargement of the gauge group from $SU(2)_{L}$ 
to $SO(4)_{L}$ which the non-linear theory does not possess. Thus, while
the actions \eq{l1} and \eq{l2} describe the same physical degrees of
freedom, the theory chooses the Lagrangian ${\cal L}$ given in \eq{l1}.

Finally, we turn to the issue of the expected duality between the
$AdS_3$ supergravity theories and two-dimensional conformal field
theories \cite{ms,m,ads1,ads2}. In \cite{ms} the authors find the states
of the conformal field theory corresponding to the representations
\eq{j0g} and \eq{j0f} originating from the coset scalars:

\be
D^{\ell+2,0)}(\ell+2,0)\;(\ell\geq-1)(0,21)\qquad \mbox{and}\qquad 
D^{\ell,0)}(\ell+4,0)(0,21)\;(\ell\geq0)
\ee

where we used labelling of Figure \ref{hotnf}. The first representation
turns out to be a chiral primary with the left and right moving weights
$h={\bar h}= (\ell+2)/2\, (\ell\ge -1)$ and the second representation a
descendant of a chiral primary with $h={\bar h}= (\ell+4)/2\, (\ell \ge
0)$; they arise in Type IIB string on $AdS_3 \times M_4$ where $M_4$ is
a $k$-fold product of $K_3$ orbifolded with the permutation group $S_k$
\cite{ms}. The resulting $N=(4,4)$ $SCFT$ has level $k$ and the
unitarity bound leads to a cutoff in the $CFT$ spectrum at $\ell=\ft12k$
not seen in the perturbation theory around $AdS_{3}$. The authors of
\cite{ms} relates this to a stringy exclusion principle for $AdS_{3}$
black holes. We expect this map to account for all the sectors of the
$AdS_{3}\times S^{3}$ spectrum presented here.

\bigskip
\noindent {\bf Acknowledgements}
\vspace{.5truecm}


We thank M. Duff, M. G\"{u}naydin, J. Maldacena and A. Strominger for useful
dicussions. We especially thank J. Maldacena for bringing the subject
matter of this paper to our attention.

\newpage


\section*{ {\large{\bf Appendix}}}


Starting from the conventions of \cite{romans1}, in which the 
signature of space-time is mostly negative, and making the replacements
\bea
g_{MN}&\rightarrow& -g_{MN}\ ,\qquad\qquad\psi_{M}\ \ \rightarrow\ \psi_{M}
\ ,\qquad
e_{M_{1}\dots M_{6}}\;\;\rightarrow\;\; -e_{M_{1}\dots M_{6}}\ ,\nn\w1
\C^{M}&\rightarrow& i\C^{M}\ ,\qquad\qquad\qquad\chi\ \ \rightarrow\  i\chi
\ ,\nn\w1
\C^{7}&\rightarrow&\C^7\ ,\qquad\qquad\qquad\quad\bar{\th}\ \ 
\rightarrow\  i\,\bar{\th}\C^{7}
\ ,\nn
\eea
we end up with a space-time with signature $(-+++++)$ and the spinor
conventions

\bea
\left\{\C^{M},\,\C^{N}\,\right\}&=& 2g^{MN} 
\ ,\qquad\qquad\qquad\qquad\ 
\bar{\th}\se i(\th)^{\dagger}\C^{0}
\ ,\nn\w2
\C^{M_{1}\dots M_{6}}&=& e^{M_{1}\dots M_{6}}\C^{7} 
\ ,\qquad\qquad\qquad
\th_{a}\se i\C^{0}(\th_{b})^{\star}\O_{ba}
\ ,\nn\w2
\ft12(1+\C^{7})\psi_{M}&=& \psi_{M}
\ ,\qquad\qquad\quad
\ft12(1-\C^{7})\chi\se \chi\ ,
\eea

where $\O$ is the anti-symmetric charge conjugation matrix of $SO(5)$,
whose $\C^{i}$-matrices are anti-symmetric. With this redefinition both
$\psi_{M}$ and $\chi$ remain symplectic Majorana-Weyl (this spinor type
is allowed in both signatures) and $C$ remains symmetric and $\C^{M}$
remain anti-symmetric. As a check one can verify the
supertransformations \eq{susy} have the correct hermicity properties
(without any factors of $i$). Both $SO(1,5)$ and $SO(5)$ spinor indices
are raised and lowered using north-east-south-west contraction. We split
the $\C^{M}$-matrices under $SO(1,5)\rightarrow SO(2,1)\times SO(3)$ as
follows

\bea
\C^{\m}&=& \c^\m \times 1\times \s_1 
\ ,\qquad\qquad\; 
C\se \e\times \y\times \s^{1}\nn\w1
\C^{a}&=& 1\times \c^a \times \s^2
\ ,\qquad\qquad
\C^{7}\se 1\times 1\times (-\s_3)\ ,
\eea

where $\e$ and $\y$ are the charge conjugation matrices. The resulting
two-component $SO(1,2)$ spinors are Majorana and the two-component
$SO(3)$ spinors are pseudo-symplectic Majorana. In our conventions
$D_{M}\e=(\del_{M}+\ft14\o_{M}{}^{AB}C_{AB})\e$ and
$[\nabla_{M},\nabla_{N}]\e=\ft14 R_{MNAB}\C^{AB}\e$.

The non-trivial part of the superalgebra $SU(1,1|2)_{L}$ is 

\be
\left\{ \,Q_{\a\,\a'\,\a''},\,Q_{\b\,\b'\,\b''}\,\right\}\se
\left(M_{\a\,\b}\,\e_{\a'\,\b'}+\e_{\a\,\b}\,T_{\a'\,\b'}\right)\e_{\a''\,\b''}
\ ,
\ee

where $\a$, $\a$, $\a''$ label the chiral spinors of $SO(2,2)$, $SO(4)$
and $SO(4)_{R}$ respectively, $M_{\a\b}=M_{\b\a}$ are the $SU(1,1)_{L}\subset SO(2,2)$
generators and $T_{\a'\,\b'}=T_{\b'\,\a'}$ are the $SU(2)_{L}\subset SO(4)$ generators.
The superalgebra $SU(1,1|2)_{R}$ has identical structure and
commutes with $SU(1,1|2)_{L}$.

\newpage


\def\ta{
{\small
\begin{table}[p]
\begin{center} 
\begin{tabular}{llcclcc} 
\hline &&&&&&\\
Helicity && \ $SO(4)$& $SO(4)_{R}\times SO(n)$ & Lowest & Conformal & Isospins\\ 
\quad $s_{0}$& &content & content &energy $E_{0}$ & weights $(h,\bar{h})$
& $(j,\bar{j})$ \\ 
&&&&&&\\ \hline &&&&&&\\
$\pm 2$ && $(\ell+1,\,0)$  &$(0,0)$ & $\ell+3$ &
\wwb{\ell+3\pm2}{\ell+3\mp2}{2} & \wwb{\ell+1}{\ell+1}{2}
\\ &&&&&&\\
$\pm 3/2$ && $(\ell+\ft32,\,\pm\ft12)$ & $(2_{\pm},0)$ & $\ell+\ft52$
& \wwb{2\ell+5\pm3}{2\ell+5\mp3}{4} & \wwb{2\ell+3\pm1}{2\ell+3\mp1}{4}
\\ &&&&&&\\
$\pm 3/2$ && $(\ell+\ft12,\,\pm\ft12)$ & $(2_{\mp},0)$ & $\ell+\ft72$
& \wwb{2\ell+7\pm3}{2\ell+7\mp3}{4} & \wwb{2\ell+1\pm1}{2\ell+1\mp1}{4}
\\ &&&&&&\\
$\pm 1$ && $(\ell+2,\,\pm1)$  &$(0,0)$ & $\ell+2$
& \wwb{\ell+2\pm1}{\ell+2\mp1}{2} & \wwb{\ell+2\pm1}{\ell+2\mp1}{2}
\\ &&&&&&\\
$\pm 1$ && $(\ell+1,\,\pm1)$ & $(4,0)$ & $\ell+3$
& \wwb{\ell+3\pm1}{\ell+3\mp1}{2} & \wwb{\ell+1\pm1}{\ell+1\mp1}{2}
\\ &&&&&&\\
$\pm 1$ && $(\ell,\,\pm1)$ & $(0,0)$ & $\ell+4$
& \wwb{\ell+4\pm1}{\ell+4\mp1}{2} & \wwb{\ell\pm1}{\ell\mp1}{2}
\\ &&&&&&\\
$\pm 1/2$ && $(\ell+\ft32,\,\pm\ft32)$ & $(2_{\mp},0)$ & $\ell+\ft52$
& \wwb{2\ell+5\pm1}{2\ell+5\mp1}{4} & \wwb{2\ell+3\pm3}{2\ell+3\mp3}{4}
\\ &&&&&&\\
$\pm 1/2$ && $(\ell+\ft12,\,\pm\ft32)$ & $(2_{\pm},0)$ & $\ell+\ft72$
& \wwb{2\ell+7\pm1}{2\ell+7\mp1}{4} & \wwb{2\ell+1\pm3}{2\ell+1\mp3}{4}
\\ &&&&&&\\
$0$ && $(\ell+1,\,\pm2)$  &$(0,0)$ & $\ell+3$
& \wwb{\ell+3}{\ell+3}{2} & \wwb{\ell+1\pm2}{\ell+1\mp2}{2}
\\ &&&&&&\\ \hline
\end{tabular} 
\end{center} 
\caption{{\small The spin $2$ tower of supermultiplets for $\ell\geq 0$.
The $SO(4)$ content is characterized by the highest weight labels
$(\ell_1,\ell_2)$. The $SO(4)_{R}\times SO(n)$ content is given by the
dimensions of the irreducible representations. The internal $SO(4)$ is
isomorphic to $SU(2)_L \times SU(2)_R$ with isospins $(j, {\bar j})$
related to the $SO(4)$ highest weight labels as $j=\ft12
(\ell_1+\ell_2)$ and ${\bar j}=\ft12 (\ell_1-\ell_2)$. Similarly
$SO(2,2) \approx SU(1,1)_L\times SU(1,1)_R$ with conformal weights
$(h,{\bar h})$ given by $h=\ft12 (E_0+s_{0})$ and ${\bar h}=\ft12
(E_0-s_{0})$. At level $\ell$ the number of bosonic and fermionic
degrees of freedom match and separately equal to $16(\ell+1)(\ell+3)$.
For the supersymmetry transformations rules see Figure \ref{httf}. } } 
\la{htt} 
\end{table}
}
}


\def\tb{
{\small
\begin{table}[p] 
\begin{center} 
\begin{tabular}{llcclcc} 
\hline &&&&&&\\
Helicity && \ $SO(4)$& $SO(4)_{R}\times SO(n)$ & Lowest& Conformal & Isospins\\ 
\quad $s_{0}$& &Content & \qquad Content &Energy $E_0$ &weights $(h,\bar{h})$
& $(j,\bar{j})$ \\ 
&&&&&&\\ \hline &&&&&&\\
$\pm 1$ && $(\ell+1,\,\mp1)$ & $(0,0)$ & $\ell+3$
& \wwb{\ell+3\pm1}{\ell+3\mp1}{2} & \wwb{\ell+1\mp1}{\ell+1\pm1}{2}
\\ &&&&&&\\
$\pm 1/2$ && $(\ell+\ft32,\,\mp\ft12)$ & $(2_{\pm},0)$ & $\ell+\ft52$
& \wwb{2\ell+5\pm1}{2\ell+5\mp1}{4} & \wwb{2\ell+3\mp1}{2\ell+3\pm1}{4} 
\\ &&&&&&\\
$\pm 1/2$ && $(\ell+\ft12,\,\mp\ft12)$ & $(2_{\mp},0)$ & $\ell+\ft72$
& \wwb{2\ell+7\pm1}{2\ell+7\mp1}{4} & \wwb{2\ell+1\mp1}{2\ell+1\pm1}{4}
\\ &&&&&&\\
$0$ && $(\ell+2,\,0)$ & $(0,0)$ & $\ell+2$
& \wwb{\ell+2}{\ell+2}{2} & \wwb{\ell+2}{\ell+2}{2}
\\ &&&&&&\\ 
$0$ && $(\ell+1,\,0)$ & $(4,0)$ & $\ell+3$
& \wwb{\ell+3}{\ell+3}{2} & \wwb{\ell+1}{\ell+1}{2}
\\ &&&&&&\\
$0$ && $(\ell,\,0)$ & $(0,0)$ & $\ell+4$
& \wwb{\ell+4}{\ell+4}{2} & \wwb{\ell}{\ell}{2}
\\ &&&&&&\\  \hline
\end{tabular} 
\end{center} 
\caption{{\small The spin $1$, $SO(n)$ singlet tower of supermultiplets for
$\ell\geq0$. At level $\ell$ the number of bosonic and fermionic degrees
of freedom match and separately equal to $8(\ell+2)^{2}$. For the supersymmetry
transformation rules see Figure \ref{hotf} and for further notation see the caption
of Table \ref{htt}.} } 
\la{hot} 
\end{table}
}
}


\def\tc{
{\small
\begin{table}[p]
\begin{center} 
\begin{tabular}{llcclcc} 
\hline &&&&&&\\
Helicity && \ $SO(4)$& $SO(4)_{R}\times SO(n)$ & Lowest& Conformal & Isospins\\ 
\quad $s_{0}$ & &Content & \qquad Content &Energy $E_{0}$ & weights $(h,\bar{h})$
& $(j,\bar{j})$ \\ 
&&&&&&\\ \hline &&&&&&\\
$\pm 1$ && $(\ell+1,\,\mp1)$  &$(0,n)$ & $\ell+3$
& \wwb{\ell+3\pm1}{\ell+3\mp1}{2} & \wwb{\ell+1\mp1}{\ell+1\pm1}{2}
\\ &&&&&&\\
$\pm 1/2$ && $(\ell+\ft32,\,\mp\ft12)$ & $(2_{\pm},n)$ & $\ell+\ft52$
& \wwb{2\ell+5\pm1}{2\ell+5\mp1}{4} & \wwb{2\ell+3\mp1}{2\ell+3\pm1}{4} 
\\ &&&&&&\\
$\pm 1/2$ && $(\ell+\ft12,\,\mp\ft12)$ & $(2_{\mp},n)$ & $\ell+\ft72$
& \wwb{2\ell+7\pm1}{2\ell+7\mp1}{4} & \wwb{2\ell+1\mp1}{2\ell+1\pm1}{4}
\\ &&&&&&\\
$0$ && $(\ell+2,\,0)$ & $(0,n)$ & $\ell+2$
& \wwb{\ell+2}{\ell+2}{2} & \wwb{\ell+2}{\ell+2}{2}
\\ &&&&&&\\ 
$0$ && $(\ell+1,\,0)$ & $(4,n)$ & $\ell+3$
& \wwb{\ell+3}{\ell+3}{2} & \wwb{\ell+1}{\ell+1}{2}
\\ &&&&&&\\
$0$ && $(\ell,\,0)$ & $(0,n)$ & $\ell+4$
& \wwb{\ell+4}{\ell+4}{2} & \wwb{\ell}{\ell}{2}
\\ &&&&&&\\ \hline
\end{tabular} 
\end{center} 
\caption{{\small The spin $1$ tower of supermultiplets in the vector
representation of $SO(n)$ for $\ell\geq -1$. At level $\ell$ the number
of bosonic and fermionic degrees of freedom match and separately equal
to $8n(\ell+2)^{2}$. For the supermultiplet transformation rules see
Figure \ref{hotnf} and for further notation see the caption of Table
\ref{htt}. }} 
\la{hotn} 
\end{table}
}
}


\def\fa{
\begin{figure}[p]
\begin{center}
\unitlength=\tsz mm
\begin{picture}(100,100)(-15,-20)
\put(50,100){\makebox(0,0){$D^{(\ell+1,0)}(\ell+3,2)(0,0)$}}
\put(35,95){\vector(-1,-1){15}}
\put(65,95){\vector(1,-1){15}}
\put(15,75){\makebox(0,0){$D^{(\ell+3/2,1/2)}(\ell+\ft52,\ft32)
(2_{+},0)$}}
\put(85,75){\makebox(0,0){$D^{(\ell+1/2,1/2)}(\ell+\ft72,\ft32)
(2_{-},0)$}}
\put(10,70){\vector(-1,-1){15}}
\put(30,70){\vector(1,-1){15}}
\put(70,70){\vector(-1,-1){15}}
\put(90,70){\vector(1,-1){15}}
\put(-10,50){\makebox(0,0){$D^{(\ell+2,1)}(\ell+2,1)(0,0)$}}
\put(50,50){\makebox(0,0){$D^{(\ell+1,1)}(\ell+3,1)(4,0)$}}
\put(110,50){\makebox(0,0){$D^{(\ell,1)}(\ell+4,1)(0,0)$}}
\put(-5,45){\vector(1,-1){15}}
\put(45,45){\vector(-1,-1){15}}
\put(55,45){\vector(1,-1){15}}
\put(105,45){\vector(-1,-1){15}}
\put(15,25){\makebox(0,0){$D^{(\ell+3/2,3/2)}(\ell+\ft52,\ft12)
(2_{-},0)$}}
\put(85,25){\makebox(0,0){$D^{(\ell+1/2,3/2)}(\ell+\ft72,\ft12)
(2_{+},0)$}}
\put(18,20){\vector(1,-1){16}}
\put(82,20){\vector(-1,-1){16}}
\put(50,0){\makebox(0,0){$D^{(\ell+1,2)}(\ell+3,0)(0,0)$}}
\put(-50,100){\makebox(0,0){$s_{0}=2$}}
\put(-50,75){\makebox(0,0){$s_{0}=\ft32$}}
\put(-50,50){\makebox(0,0){$s_{0}=1$}}
\put(-50,25){\makebox(0,0){$s_{0}=\ft12$}}
\put(-50,0){\makebox(0,0){$s_{0}=0$}}
\end{picture}
\end{center}
\caption{{\small The spin $2$ supermultiplet structure for $\ell\geq0$.
A given representation is denoted by $ D^{(\ell_1,\ell_2)}\ (E_0,s_{0})\
(R\times S\,)$ where $(\ell_1,\ell_2)$ label the $S^3$ isometry group
$SO(4)$; $(E_0,s_{0})$ label the representation of the AdS group
$SO(2,2)$; $R$ denotes the representation of $SO(4)_R$ and $S$ denotes a
representation of $SO(n)$. The supercharge
$Q_{+}^{(1/2,\,1/2)}(-\ft12,\,-\ft12)(2_{+},0)$ acts to the southwest
and the supercharge $Q_{-}^{(1/2,\,-1/2)}(\ft12,\,-\ft12)(2_{-},0)$ acts
to the southeast. The full multiplet structure is obtained by adding the
conjugate tower of multiplets in which the replacements
$\ell_{2}\rightarrow -\ell_{2}$, $s_{0}\rightarrow -s_{0}$,
$2_{\pm}\rightarrow 2_{\mp}$ are made. The multiplet at level $\ell$
thus contains $16(\ell+1)(\ell+3)$ Bose and that many Fermi states. At
level $\ell=-1$ there are $SU(2)_{L}$ Yang-Mills states with
$(E_{0},\,s_{0})=(1,1)$ and $SU(2)_{R}$ Yang-Mills states with
$(E_{0},\,s_{0})=(1,-1)$ which are pure gauge. At level $-1$ one also
finds the non-propagating graviton and gravitini. At level $\ell=0$, the
states on the lower right diagonal are absent and the resulting special
spin $2$ multiplet consists of 48 Bose and 48 Fermi degrees of freedom.
At level $\ell=1$ there are physical $SU(2)_{L}$ Yang-Mills states with
$(E_{0},\,s_{0})=(5,1)$ and physical $SU(2)_{R}$ Yang-Mills states with
$(E_{0},\,s_{0})=(5,-1)$. For $\ell\geq 1$ the structure of the
multiplets is generic, and no other Yang-Mills states arise.} }
\la{httf}
\end{figure}
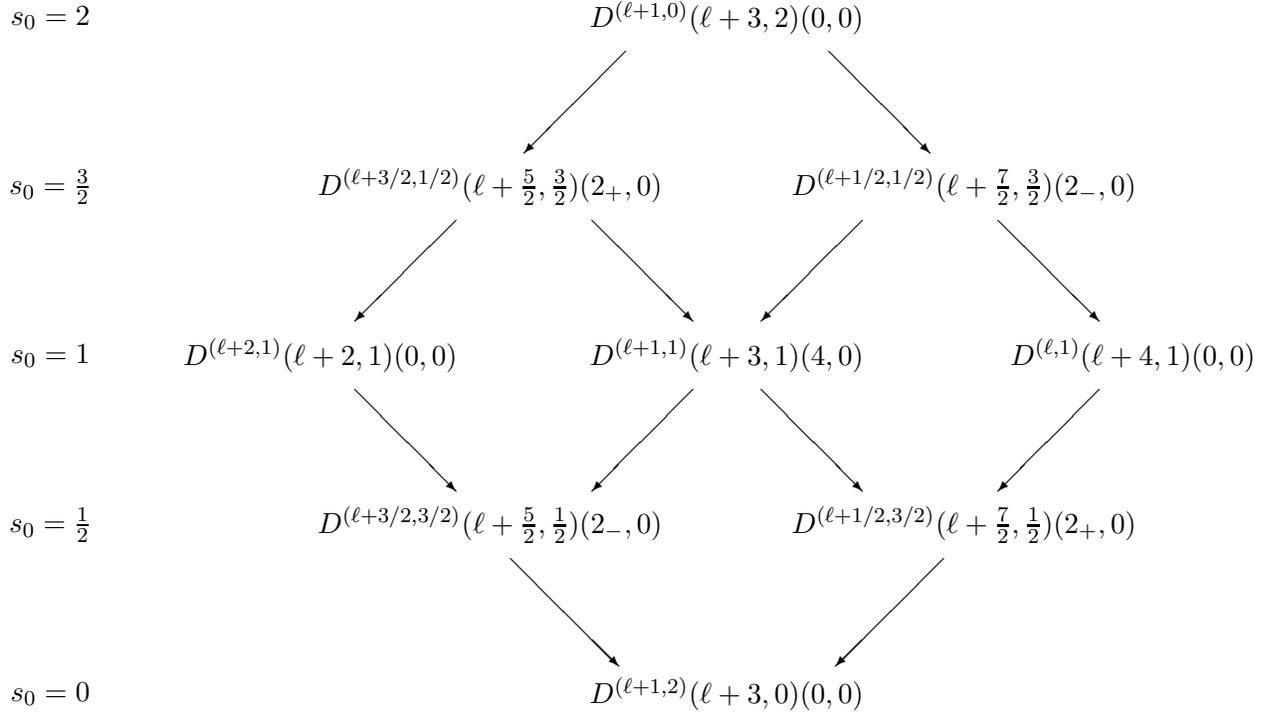
}


\def\fb{
\begin{figure}[h]
\begin{center}
\unitlength=\tsz mm
\begin{picture}(100,100)(-15,-20)
\put(50,100){\makebox(0,0){$D^{(\ell+1,-1)}(\ell+3,1)(0,0)$}}
\put(35,95){\vector(-1,-1){15}}
\put(65,95){\vector(1,-1){15}}
\put(15,75){\makebox(0,0){$D^{(\ell+3/2,-1/2)}(\ell+\ft52,\ft12)
(2_{+},0)$}}
\put(85,75){\makebox(0,0){$D^{(\ell+1/2,-1/2)}(\ell+\ft72,\ft12)
(2_{-},0)$}}
\put(10,70){\vector(-1,-1){15}}
\put(30,70){\vector(1,-1){15}}
\put(70,70){\vector(-1,-1){15}}
\put(90,70){\vector(1,-1){15}}
\put(-10,50){\makebox(0,0){$D^{(\ell+2,0)}(\ell+2,0)(0,0)$}}
\put(50,50){\makebox(0,0){$D^{(\ell+1,0)}(\ell+3,0)(4,0)$}}
\put(110,50){\makebox(0,0){$D^{(\ell,0)}(\ell+4,0)(0,0)$}}
\put(-5,45){\vector(1,-1){15}}
\put(45,45){\vector(-1,-1){15}}
\put(55,45){\vector(1,-1){15}}
\put(105,45){\vector(-1,-1){15}}
\put(15,25){\makebox(0,0){$D^{(\ell+3/2,1/2)}(\ell+\ft52,-\ft12)
(2_{-},0)$}}
\put(85,25){\makebox(0,0){$D^{(\ell+1/2,1/2)}(\ell+\ft72,-\ft12)
(2_{+},0)$}}
\put(18,20){\vector(1,-1){16}}
\put(82,20){\vector(-1,-1){16}}
\put(50,0){\makebox(0,0){$D^{(\ell+1,1)}(\ell+3,-1)(0,0)$}}
\put(-50,100){\makebox(0,0){$s_{0}=1$}}
\put(-50,75){\makebox(0,0){$s_{0}=\ft12$}}
\put(-50,50){\makebox(0,0){$s_{0}=0$}}
\put(-50,25){\makebox(0,0){$s_{0}=-\ft12$}}
\put(-50,0){\makebox(0,0){$s_{0}=-1$}}
\end{picture}
\end{center}
\caption{{\small The spin $1$, $SO(n)$ singlet supermultiplet structure
for $\ell\geq0$. The multiplet is {\it self-conjugate} and contains
$8(\ell+2)^{2}$ Bose and that many Fermi states at level $\ell$. At
level $\ell=-1$ there is an unphysical spin $\ft12$ multiplet residing
at the left diamond. At level $\ell=0$ there is triplet of $SU(2)_{R}$
Yang-Mills states with $(E_{0},\,s_{0})=(3,1)$ and a triplet of
$SU(2)_{L}$ Yang-Mills states with $(E_{0},\,s_{0})=(3,-1)$. For
$\ell\geq 0$ the structure of the multiplets is generic and no other
Yang-Mills states arise. See caption of Figure \ref{httf} for furher
notations. }}
\la{hotf}
\end{figure}
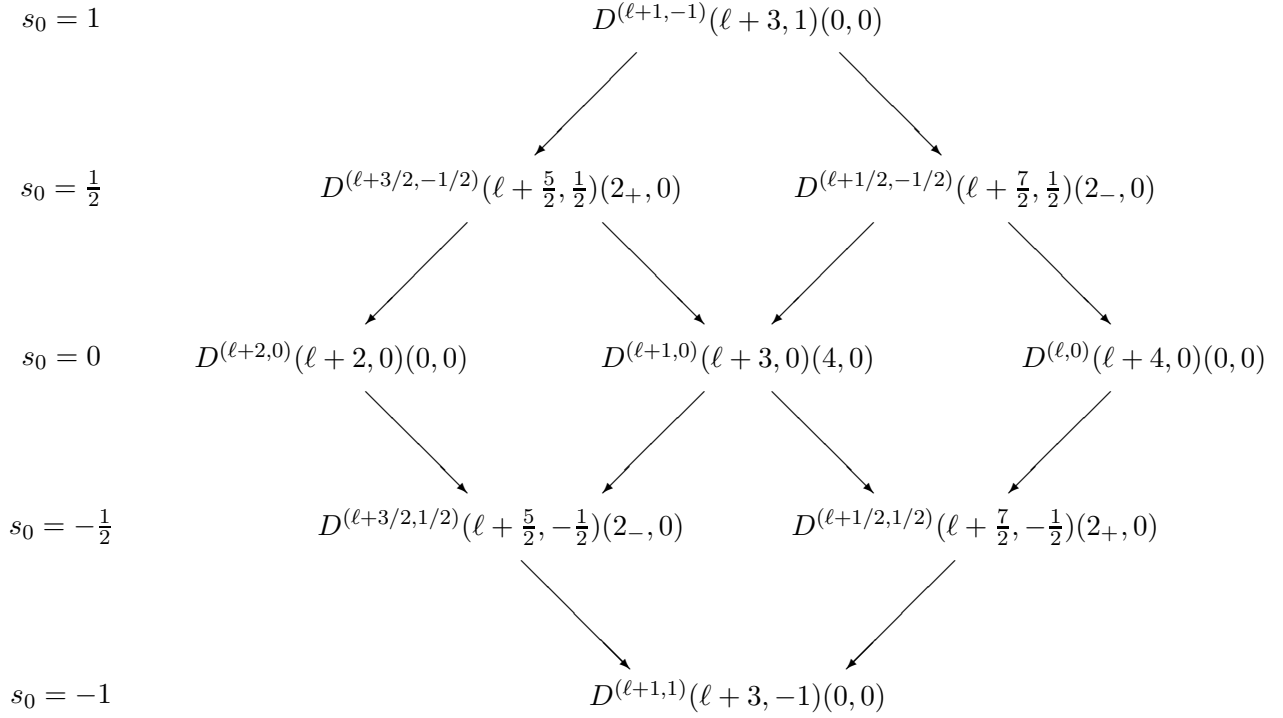
}


\def\fc{
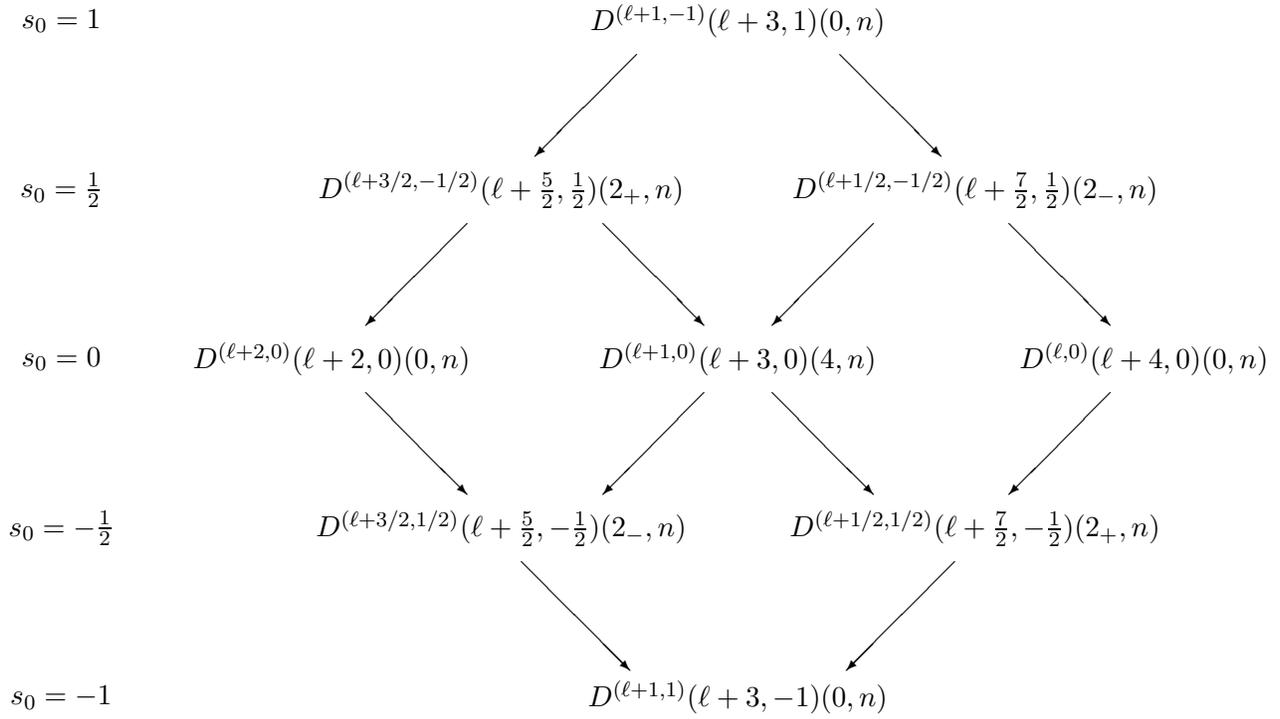
\begin{figure}[h]
\begin{center}
\unitlength=\tsz mm
\begin{picture}(100,100)(-15,-20)
\put(50,100){\makebox(0,0){$D^{(\ell+1,-1)}(\ell+3,1)(0,n)$}}
\put(35,95){\vector(-1,-1){15}}
\put(65,95){\vector(1,-1){15}}
\put(15,75){\makebox(0,0){$D^{(\ell+3/2,-1/2)}(\ell+\ft52,\ft12)
(2_{+},n)$}}
\put(85,75){\makebox(0,0){$D^{(\ell+1/2,-1/2)}(\ell+\ft72,\ft12)
(2_{-},n)$}}
\put(10,70){\vector(-1,-1){15}}
\put(30,70){\vector(1,-1){15}}
\put(70,70){\vector(-1,-1){15}}
\put(90,70){\vector(1,-1){15}}
\put(-10,50){\makebox(0,0){$D^{(\ell+2,0)}(\ell+2,0)(0,n)$}}
\put(50,50){\makebox(0,0){$D^{(\ell+1,0)}(\ell+3,0)(4,n)$}}
\put(110,50){\makebox(0,0){$D^{(\ell,0)}(\ell+4,0)(0,n)$}}
\put(-5,45){\vector(1,-1){15}}
\put(45,45){\vector(-1,-1){15}}
\put(55,45){\vector(1,-1){15}}
\put(105,45){\vector(-1,-1){15}}
\put(15,25){\makebox(0,0){$D^{(\ell+3/2,1/2)}(\ell+\ft52,-\ft12)
(2_{-},n)$}}
\put(85,25){\makebox(0,0){$D^{(\ell+1/2,1/2)}(\ell+\ft72,-\ft12)
(2_{+},n)$}}
\put(18,20){\vector(1,-1){16}}
\put(82,20){\vector(-1,-1){16}}
\put(50,0){\makebox(0,0){$D^{(\ell+1,1)}(\ell+3,-1)(0,n)$}}
\put(-50,100){\makebox(0,0){$s_{0}=1$}}
\put(-50,75){\makebox(0,0){$s_{0}=\ft12$}}
\put(-50,50){\makebox(0,0){$s_{0}=0$}}
\put(-50,25){\makebox(0,0){$s_{0}=-\ft12$}}
\put(-50,0){\makebox(0,0){$s_{0}=-1$}}
\end{picture}
\end{center}
\caption{{\small The structure of the spin $1$ supermultiplet in the
vector representation of $SO(n)$ for $\ell\geq-1$. The multiplet is {\it
self-conjugate} and contains $8\,n\,(\ell+2)^2$ Bose and that many
Fermi states at level $\ell$. For the special value $\ell=-1$ one finds
a spin $\ft12$ multiplet consisting of $8\,n$ Bose and $8\,n$ Fermi
states residing at the left diamond. For $\ell\geq 0$ the structure of
the multiplets is generic. These are matter spin $1$ multiplets. See the
caption of Figure \ref{httf} for further notations.}}
\la{hotnf}
\end{figure}
}


\def\bbl{

}

\bbl
\newpage
\ta
\newpage
\tb
\tc
\newpage
\fa
\newpage
\fb
\newpage
\fc

\end{document}